\documentclass[superscriptaddress,aps,pre,twocolumn]{revtex4-2}

\usepackage[sort&compress]{natbib}
\usepackage{amsmath}
\usepackage{amsfonts}
\usepackage{amssymb}
\usepackage{mathtools} % for "\newtagform" macro
\newtagform{brackets}{[}{]}
\usetagform{brackets} % employ square brackets as delimiters around eq. numbers

\usepackage[colorlinks]{hyperref}

\hypersetup{colorlinks=true,
	linkcolor=blue,
	anchorcolor=black,
	citecolor=red,
	urlcolor=black
}

\usepackage[capitalise]{cleveref}
\creflabelformat{equation}{#2{\upshape[#1]}#3}

\usepackage[font=small,labelfont=bf,justification=justified,format=plain]{caption}
\usepackage{subfig}

\newcommand{\figCaption}[2]{\caption[#1]{\textbf{#1}. #2}}

\providecommand{\smdna}{{\rm \scriptscriptstyle DNA}}
\providecommand{\smdnas}{{\rm \scriptscriptstyle D}}
\providecommand{\smf}{{\rm \scriptscriptstyle F}}

\providecommand{\be}{\begin{equation}}
\providecommand{\ee}{\end{equation}}
\providecommand{\bea}{\begin{eqnarray}}
\providecommand{\eea}{\end{eqnarray}}
\providecommand{\beas}{\begin{eqnarray*}}
\providecommand{\eeas}{\end{eqnarray*}}
\providecommand{\bal}{\begin{align}}
\providecommand{\eal}{\end{align}}
\providecommand{\nn}{\nonumber}

\begin{document}

\title{Steric interactions and out-of-equilibrium processes control the internal organization of bacteria}

% Please add a significance statement to explain the relevance of your work
%\significancestatement{Self-organizing processes and structures lie at the heart of biology. Here, we study compaction and localization of DNA inside cells of the bacterium \textit{Escherichia coli}. We explain the physics behind this phenomenon based on excluded-volume interactions, by which the DNA spontaneously phase separates from other intracellular crowders. We then explore the rich phenomenology that arises when these interactions are coupled to out-of-equilibrium cellular processes, such as the transcription and degradation of messenger RNAs. The results agree well with recent experimental observations of the size, positioning, and dynamical splitting of \textit{E. coli}’s DNA, providing an example of the tight control that cells exert over their internal organization by means of out-of-equilibrium processes.}

\begin{abstract}
Despite the absence of a membrane-enclosed nucleus, the bacterial DNA is typically condensed into a compact body---the nucleoid. This compaction influences the localization and dynamics of many cellular processes including transcription, translation, and cell division. Here, we develop a model that takes into account steric interactions among the components of the \textit{Escherichia coli} transcriptional-translational machinery (TTM) and out-of-equilibrium effects of mRNA transcription, translation, and degradation, in order to explain many observed features of the nucleoid. We show that steric effects, due to the different molecular shapes of the TTM components, are sufficient to drive equilibrium phase separation of the DNA, explaining the formation and size of the nucleoid. In addition, we show that the observed positioning of the nucleoid at midcell is due to the out-of-equilibrium process of messenger RNA (mRNA) synthesis and degradation: mRNAs apply a pressure on both sides of the nucleoid, localizing it to midcell. We demonstrate that, as the cell grows, the production of these mRNAs is responsible for the nucleoid splitting into two lobes, and for their well-known positioning to $1/4$ and $3/4$ positions on the long cell axis. Finally, our model quantitatively accounts for the observed expansion of the nucleoid when the pool of cytoplasmic mRNAs is depleted. Overall, our study suggests that steric interactions and out-of-equilibrium effects of the TTM are key drivers of the internal spatial organization of bacterial cells.
\end{abstract}

% Use letters for affiliations, numbers to show equal authorship (if applicable) and to indicate the corresponding author
\author{Ander Movilla Miangolarra}
\affiliation{Laboratoire Physico-Chimie Curie, Institut Curie, PSL Research University, CNRS UMR 168, Paris, France}
\affiliation{Sorbonne Universit\'es, UPMC Univ. Paris 06, Paris, France} 
\author{Sophia Hsin-Jung Li}
\affiliation{Department of Molecular Biology, Princeton University, Princeton, NJ 08544, United States}
\author{Jean-Fran\c{c}ois Joanny}
\affiliation{Laboratoire Physico-Chimie Curie, Institut Curie, PSL Research University, CNRS UMR 168, Paris, France}
\affiliation{Sorbonne Universit\'es, UPMC Univ. Paris 06, Paris, France} 
\affiliation{ESPCI Paris, PSL Research University, 10 rue Vauquelin, 75005, Paris, France}
\affiliation{Collège de France, 11 Place Marcelin Berthelot, 75231, Paris Cedex05, France}
\author{Ned S. Wingreen}
\altaffiliation{To whom correspondence should be addressed. E-mails: wingreen@princeton.edu and michele.castellana@curie.fr}
\affiliation{Department of Molecular Biology, Princeton University, Princeton, NJ 08544, United States}
\affiliation{Lewis-Sigler Institute for Integrative Genomics, Princeton University, Princeton, NJ 08544, United States}
\author{Michele Castellana}
\altaffiliation{To whom correspondence should be addressed. E-mails: wingreen@princeton.edu and michele.castellana@curie.fr}
\affiliation{Laboratoire Physico-Chimie Curie, Institut Curie, PSL Research University, CNRS UMR 168, Paris, France}
\affiliation{Sorbonne Universit\'es, UPMC Univ. Paris 06, Paris, France} 

\date{\today}

\maketitle

Living systems show a high degree of organization at multiple scales, from the molecular scale to the macroscopic scales of organisms and ecosystems. A notable example of spatial organization in cells is that of bacterial DNA: despite the absence of a nuclear membrane, in many bacteria such as \textit{E. coli}, the chromosome is  not randomly spread throughout the intracellular space, but is markedly localized \cite{Lewis2004Bacterial,bakshi2012superresolution}, and forms a compact structure---the nucleoid. The localization and degree of confinement of the nucleoid varies with growth rate and among bacterial species \cite{gray2019nucleoid}. This organization and localization of the chromosome has been shown to play an important role in many biological processes, including transcription via the distribution of RNA polymerases \cite{Weng2019SpatialRNAP}, translation via localization of ribosomes \cite{Sanamrad2014TrackingRibosomes}, and the localization and diffusion of protein aggregates \cite{Coquel2013LocalizationProtein}.

Despite the importance of chromosome localization, the physical causes and regulatory mechanisms of its confinement are still largely unknown. One of the causes of the compaction of the nucleoid could 
be the fact that the cytoplasm acts as a poor solvent for the chromosome \cite{Xiang2020SolventQuality}, but many other factors could also affect nucleoid compaction, such as nucleoid-associated proteins that modify the folding conformation of the chromosome \cite{Dame2020}. Regarding localization, various studies have shown that, while the nucleoid is located at the center of the cell before chromosome replication, during and after replication the daughter chromosomes move out of the center \cite{joshi2011escherichia}, typically localizing at $1/4$ and $3/4$ positions on the long cell axis \cite{wu2019cell}. With respect to nucleoid size, one of the determinants could be macromolecular crowding \citep{Cabrera2009ActiveTranscription}.

Previous theoretical efforts to explain the compaction and localization of the nucleoid have been mostly based on Monte Carlo simulations \cite{joyeux_microorganism,mondal2011entropy}. It was found that excluded-volume effects between DNA and polysomes---messenger RNAs (mRNAs) bound to multiple ribosomes---may account for segregation of the nucleoid from the rest of the cytoplasm \cite{mondal2011entropy}. However, the mechanism for the positioning of the nucleoid, both before and after chromosome replication, remains unclear, and  has been hypothesized to require an active process \cite{joyeux_microorganism}.

In this study, we develop a statistical-physics description of the spatial localization of the molecular components of the \textit{E. coli} transcriptional-translational machinery (TTM)---composed of DNA, mRNAs, and ribosomes---and identify the physical mechanisms underlying their localization patterns. Unlike previous studies, we leverage semi-analytical methods, e.g.,  the virial expansion, which allow us to tackle the complexity of the system, and reduce it to a set of computationally tractable reaction-diffusion equations. In previous approaches the localization of the nucleoid was imposed as an input of the model \cite{castellana2016spatial}. By contrast, here we quantitatively demonstrate that localization patterns on the cellular scale spontaneously emerge from microscopic features on a molecular scale, i.e., steric effects between DNA and polysomes. 
In particular, we show that the segregation of the nucleoid from mRNAs and ribosomes is caused by equilibrium excluded-volume effects only, as in classical phase separation. Also, our analysis shows that other dynamical features, such as nucleoid positioning at midcell and at $1/4$ and $3/4$ along the cell axis, are driven by the synthesis and degradation of mRNAs, making it a purely out-of-equilibrium feature. Finally, we compare these results with experimental data obtained from cells growing filamentously \cite{wu2019cell}, either with or without chromosome replication, providing important physical and mechanistic insights.

\begin{figure*}%[tbhp]
\centering
{\includegraphics[width=\textwidth]{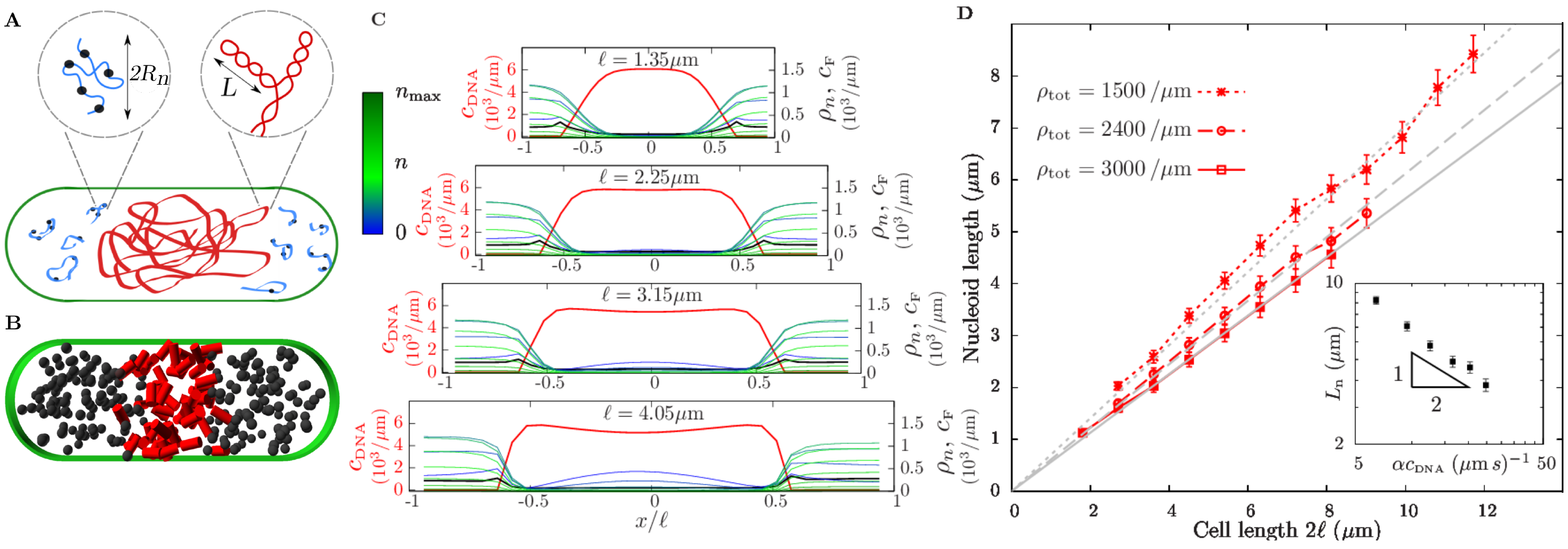}}
\figCaption{Steady-state concentration profiles for \textit{E. coli}  growing filamentously}{\label{fig1}
(\textit{A}) Cartoon of  an \textit{E. coli} cell and its transcriptional-translational machinery, where the horizontal axis is the single dimension we consider. Blue  coils represent  mRNAs in polysomes, the red coil denotes the DNA plectoneme, and ribosomes are shown in black. Blow-ups: polysome composed of an mRNA and $n$ ribosomes with gyration radius $R_n$ (left) and plectonemic structure of the DNA with persistence length $L$ (right).
(\textit{B}) Schematic of the components underlying the reaction-diffusion model, where the DNA plectoneme is represented by a set of disjoint cylinders, and polysomes and free ribosomes by spheres.
(\textit{C}) Concentrations along the long axis of the cell of DNA, $c_{\smdna}(x)$ (red), and free ribosomes, $c_{\smf}(x)$ (black), and polysomes $\rho_n(x)$ with the mRNA loading number $n$ indicated by the color bar. Each panel corresponds to a different cell half-length $\ell$, marked on the top of the panel, with total mRNA density $\rho_{\rm tot}=2400 \mu \textrm{m}^{-1}$. (\textit{D}) Nucleoid length versus cell length, $2\ell$, for different values of the total mRNA density, $\rho_{\rm tot}$. For each value of $\rho_{\rm tot}$ the nucleoid length is shown up to the cell length at which the nucleoid  splits into two lobes. The inset depicts on a log-log plot the power-law relationship between the 
length at which the nucleoid splits and the rate of mRNA synthesis $\alpha c_\smdna$, where $c_\smdna$ is the average DNA concentration along the nucleoid at steady state.}
\end{figure*}

\section*{Results}\label{results}

We describe the dynamics of the \textit{E. coli} TTM by means of a minimal out-of-equilibrium statistical-physics model. We first derive dynamical equations for the currents of DNA segments, mRNAs and ribosomes, incorporating steric effects and using the virial expansion to compute the local free energy.
This procedure is illustrated for a toy system of a binary mixture of hard spheres in \textit{SI Appendix}, Section  \ref{toy_model}, and then for the full TTM in Section \ref{currents_model}. 

By observing that \textit{E. coli} cells have an approximately cylindrical shape and symmetry, we reduce the three-dimensional cytoplasm to a single dimension 
along the long cell axis (see Fig. \ref{fig1}A and B) and describe the TTM in terms of the one-dimensional concentrations of DNA segments, mRNAs, and ribosomes. Namely, we denote by $c_{\smdna}(x,t)$  the concentration of DNA plectoneme segments  at position $x$ along the long cell axis and time $t$, by $\rho_n(x,t)$ that of polysomes composed of an mRNA and $n$ ribosomes, and by $c_{\smf}(x,t)$ that of freely diffusing ribosomes, see Fig. \ref{fig1}A and B. 
We then consider the reaction-diffusion equations for these concentrations, where we incorporate the currents and the chemical reactions, i.e., ribosome-mRNA binding and unbinding, mRNA synthesis and degradation:
\begin{align}
\partial_t c_{\smdna}(x,t) =& -\partial_x J_{\smdna} (x,t), \label{RD_eqs_C}\\
\partial_t \rho_n (x,t) =& -\partial_x J_n (x,t) - k_{\textrm{on}} c_{\smf} (x,t) \rho_{n}(x,t) \nonumber \\ &-k_{\textrm{off}} \, n \, \rho_n(x,t)  +  k_{\textrm{on}} c_{\smf} (x,t) \rho_{n-1}(x,t)  \nn \\ 
 &+ k_{\textrm{off}} (n+1)\rho_{n+1}(x,t) + \alpha\, c_{\smdna}(x,t) \delta_{n,0} \nonumber \\ & -\beta \rho_{n} (x,t), \label{RD_eqs_n} \\  \label{reac_diff}
\partial_t c_{\smf} (x,t) =& -\partial_x J_{\smf}(x,t)-k_{\textrm{on}} c_{\smf}(x,t)\sum_{n} \rho_n(x,t) \nonumber \\ & 
+ k_{\textrm{off}}\sum_{n} n\rho_n(x,t) + \beta\sum_{n} n\rho_n(x,t).
\end{align}
\noindent
In \crefrange{RD_eqs_C}{reac_diff}, $J_{\smdna}$, $J_n$, and $J_{\smf}$  denote the particle currents (derived in \textit{SI Appendix}, Sections  \ref{toy_model} and \ref{currents_model}), $k_{\textrm{on}}$ and $k_{\textrm{off}}$ the rate constants for ribosome binding and unbinding due to completion of translation, respectively, $\alpha$  the rate at which mRNAs are created locally by transcription, and $\beta$ the mRNA degradation rate.

Regarding the steric interactions, as shown in Fig. \ref{fig1}A and B, we consider ribosomes as spheres of radius $R$ and, because mRNAs and polysomes with $n$ ribosomes are globular polymer coils, we also approximate them  as spheres of radius $R$ and $R_n$, respectively. Because the \textit{E. coli} DNA has a branched, plectonemic structure with a well-defined persistence length and transverse radius  \cite{odijk2000dynamics}, we consider the chromosome as a set of cylindrical segments, where the length of each segment corresponds to the persistence length. For the sake of computational tractability, we treat the  DNA segments as disconnected as in Fig. \ref{fig1}B. 

When deriving the particle currents, the quantity of interest is the free energy of the particles. For independent spherical particles (ribosomes and polysomes), the entropic term  in the free energy is included in the virial expansion. However, for the DNA plectoneme the situation is different, 
due to the connectivity between the DNA ``cylinders''. While connectivity should not have a large effect on the virial terms, it is not clear that this is the case for entropy. Nevertheless, we find that  this entropic term is small compared to the virial terms of the DNA free energy and therefore we neglect it (see \textit{SI Appendix}, Section \ref{DNA_entropy}).

\subsection*{Model parameters}
We fix the model parameters from experiments as follows. First, we consider the parameters on a molecular scale: The radius and length of DNA cylinders are $\rho = 10 \, \rm nm$ and $L = 200 \, \rm nm$ \cite{mondal2011entropy,odijk2000dynamics}, respectively, where $L/2$ is approximately the persistence length of a DNA plectoneme \cite{cunha2001polymer,odijk2000dynamics}. However, two overlapping DNA plectonemes may be nested into each other, as discussed in \cite{mondal2011entropy}. To model this nesting, while we  use the radius $\rho$ to describe overlaps between a DNA cylinder and ribosomes or mRNAs in the virial expansion, we use a smaller, effective radius $\rho'<\rho$ for overlaps between two DNA cylinders  \cite{mondal2011entropy}, see \textit{SI Appendix}, Section  \ref{currents_model} for details.

We take the ribosome radius to be $R = 10 \, \rm nm$ \cite{mondal2011entropy}, and the radius of a ribosome-free mRNA to be $R_0 = 20 \, \rm nm$ \cite{kaczanowska2007ribosome}. The radius $R_n$ of an mRNA loaded with $n$ ribosomes is estimated  as the sum of the volume of a bare mRNA and $n$ times the volume of a ribosome, i.e., $4/3 \pi(R_0^3+n R^3)$ yielding $R_n = (R_0^3+n R^3)^{1/3}$.

We estimated the diffusion constant of the different species as follows: $D_{\smf} = 0.4 \, \mu \rm m^2/s$ for ribosomes, and $D_n = 5 \times 10^{-2} \mu \rm m^2/s$ for bare mRNAs and polysomes \cite{bakshi2012superresolution,sanamrad2014single}. Because DNA segments have a linear dimension similar to that of polysomes,  we assume that their diffusion coefficients will also be similar and take   $D_{\smdna} = 10^{-2} \, \mu \rm m^2/s$. 

The parameters relative to the cellular scale are the total number of ribosomes per cell $N_{\smf}$, the cell half-length $\ell$, both of which will be varied, and the radius of the cellular cross section, which is held constant.
%, corresponding to a reference cell with a doubling time of $\sim 2 \, \rm hr$ and length $2 \ell \sim 1.8 \, \mu \rm m$. 
Because a central aim is to compare to the experiments in Ref. \cite{wu2019cell}, we are interested in values for a doubling time of $\sim 2 \, \rm hr$ as in that study.
We  thus interpolated  experimental data points for different growth rates, to obtain the parameter values for the desired growth rate (see \textit{SI Appendix}, Sections  \ref{experiments} and  \ref{parameters_est}) and obtained a total number  $N_{\smf}\sim 7300$ ribosomes, a cross-sectional radius $R_{\rm cell} \approx 0.4 \, \rm \mu m,$ and a  cell half-length $\ell \sim 0.9 \, \mu \rm m$ for a reference cell. 
%In this case we estimated the number of ribosomes from experimental measurements of rRNA, assuming that it accounts for two thirds of the total mass of ribosomes \ref{biochemistry_berg}. 
In addition, the total mRNA concentration  for the reference cell  was fixed at 
$\rho_{\rm tot} = \sum_n \rho_n = 2400 \, \mu \textrm{m}^{-3}$ \cite{bartholomaus2016mRNA}.
The total number of DNA cylinders for the reference cell was taken to be $N_{\smdna} \sim 6700$ segments \cite{mondal2011entropy}. 

Finally, we set the reaction rates to $k_{\rm on} = 6\times 10^{-4} \mu \rm m/s$, $k_{\rm off} = 2.5\times 10^{-2} \rm /s$ \cite{castellana2016spatial}, the mRNA degradation rate $\beta = 3 \times 10^{-3}/\rm s$ corresponds to an mRNA half life of $\sim 5\,  \rm min$ \cite{bernstein2004global}, and the mRNA synthesis rate $\alpha$ is estimated from the global steady-state condition of  \cref{RD_eqs_n},
$\alpha N_{\rm DNA} = \beta N_{\rm mRNA}$ \cite{castellana2016spatial}, where $N_{\rm mRNA}$ is the total number of mRNA molecules in the cell, i.e., $\rho_{\rm tot}$ times the cell volume. 

In order to understand the compaction and localization of the bacterial nucleoid, we solved the  one-dimensional reaction-diffusion \crefrange{RD_eqs_C}{reac_diff}. To compare the predictions of our model with experimental data, in what follows, we consider two scenarios for how the concentrations of the molecular species scale with cell length.

\begin{figure}
\includegraphics[scale=1.44]{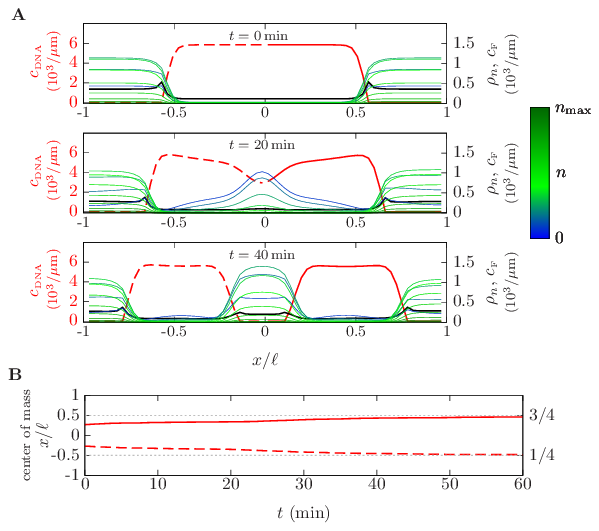}
\figCaption{Out-of-equilibrium mechanisms  split the nucleoid into two lobes located at 1/4 and 3/4 of the cell length} {\label{fig3} (\textit{A}) Concentration profiles for a filamentous cell, obtained from the equilibrium profile at $t=0$ by integrating forward in time the reaction-diffusion \crefrange{RD_eqs_C}{reac_diff} in the presence of the out-of-equilibrium processes until $t =20 \, \rm min$ and $t =40 \, \rm min$, for a cell with a half-length  $ \ell  = 4.05 \, \mu \rm m$ shown in the lowest panel of Fig. \ref{fig1}C. (\textit{B}) Positions of the center of mass of the left (dashed red curve) and right (red curve) halves of the DNA along the long axis of the cell as fraction of the total cell length, as functions of time.}
\end{figure}

\subsection*{Filamentous growth}

In filamentous growth, the total number of DNA segments, mRNAs, and ribosomes is  proportional to the cell length. For each cell length, we first determined the equilibrium steady state of the system by minimizing the free energy (\textit{SI Appendix}, \cref{F_tot}), and then numerically integrated the reaction-diffusion  \crefrange{RD_eqs_C}{reac_diff} forward in time to reach an out-of-equilibrium steady state---see \textit{SI Appendix}, Section  \ref{numerics} for details. The minimization of the free energy takes into account the steric interactions between particles and predicts the existence of a phase-separated nucleoid in the cell. The out-of-equilibrium steady state is obtained by switching on the chemical reactions. The results are shown in Fig. \ref{fig1}C and D for different cell lengths, up to the cell length at which the nucleoid spontaneously splits into two lobes, and for different total mRNA densities. 

The relation between the nucleoid length and cell length appears to be roughly linear up until the cell length at which the nucleoid begins to split in two, as seen in Fig. \ref{fig1}C and D. In our model, the nucleoid size (provided the nucleoid is single lobed) is mainly set by the balance of  osmotic pressures between the nucleoid and the peripheral cytoplasm. These pressures solely stem from the entropy and steric interactions of the components of the mixture, making the nucleoid size a consequence of equilibrium physics: in fact, in  \textit{SI Appendix}, Fig. \ref{equil_const_crowd} we show the steady-state profile of the system in the absence of out-of-equilibrium terms, displaying a linear relation between nucleoid size and cell size analogous to that of the nonequilibrium case. Moreover, as shown in Fig. \ref{fig1}D, the higher the total mRNA density, the smaller the nucleoid, implying that a high mRNA density increases the osmotic pressure on the nucleoid, thus making it shrink. 

The dependence of the nucleoid size on the cell length can be quite accurately understood from a simplified model as follows. Consider the cell as a cylindrical container, divided in three parts by two movable walls. The chromosome is confined in the central container, while mRNAs and ribosomes are equally divided in the flanking ones. The walls will reach an equilibrium position at the point where the osmotic pressures between the compartments are balanced. In view of the steric interactions in each compartment, the osmotic pressure exerted by the $i$th compartment, to first order in the virial expansion (see \textit{SI Appendix}, Section  \ref{estimates}), can be written as:
\begin{equation}
\label{pressures}
P_i=\frac{k_B T N_i}{V_i}\left(1+\frac{N_iB_i}{2V_i}\right),
\end{equation}
where $N_i$ is the number of particles in compartment $i$, $V_i=\sigma L_i$ the volume of the compartment, $\sigma$ the cross-section of the cell cylinder, $L_i$ the compartment length, and $B_i$ the pairwise virial coefficient associated with the interaction among particles. In the central compartment,  $B_i$ corresponds to the virial coefficient between DNA segments, while for the flanking compartments we  take $B_i$ as an effective virial coefficient, obtained by assuming that all ribosomes are bound to mRNAs, and equally distributed among them. By equating the pressures of the different compartments, we obtain the gray lines in Fig. \ref{fig1}D---see  \textit{SI Appendix}, Section  \ref{estimates} for details.

\begin{figure*}
\centering
\includegraphics[width=0.75\textwidth]{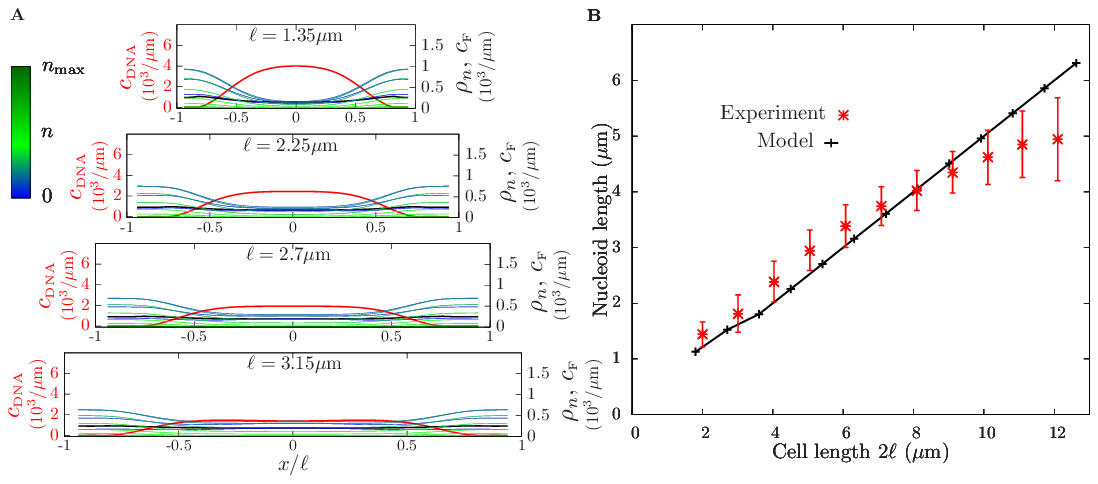}
\figCaption{Steady-state concentration profiles, for single-chromosome filamentous growth}{
\label{fig2}(\textit{A}) Concentrations of DNA, $c_{\smdna}(x)$, free  ribosomes $c_{\rm F}(x)$,  and polysomes $\rho_n(x)$, shown as in Fig. \ref{fig1}C. (\textit{B}) Nucleoid length and standard deviation as a function of cell length from Ref. \cite{wu2019cell} (red) and from the model (black).}
\end{figure*}

In \cref{pressures}, the two terms in $1+N_iB_i/(2V_i)$ encode the two different factors that contribute to the osmotic pressure. The first term stems from the entropic pressure of an ideal gas while the second one, $N_iB_i/V_i$, comes from the inter-particle interactions. In the parameter range of Fig. \ref{fig1}D, the interaction term is typically between $10\%$ and $20\%$ larger than the entropic one (however, the effect of introducing a third-order virial coefficient is small, it is of the order of a tenth of the entropic term). We find that the inclusion of steric terms in both the nucleoid and mRNA/ribosome compartments makes the nucleoid swell compared to what its size would be with only entropic terms (ideal gas contribution). This is due to the nature of the nucleoid, a long relatively stiff polymer with little entropy per segment compared to all ribosomes and mRNAs collectively.

While the linear increase of nucleoid length with cell length is the result of equilibrium osmotic pressure balance, the splitting of the nucleoid is entirely due to out-of-equilibrium processes. In fact, for cells with $\rho_\textrm{tot}=2400/\mu \textrm{m}$ and a half length of $\sim 4\, \mu {\rm m}$ or larger, the equilibrium steady state used as the initial condition for the reaction-diffusion  equations yields a nucleoid with a single lobe. By contrast, the nucleoid splits into two identical lobes positioned at $1/4$ and $3/4$ of the long cell axis when the reaction-diffusion \crefrange{RD_eqs_C}{reac_diff} are integrated forward in time, see Fig. \ref{fig3} and \textit{Movie S1}. Such a $1/4$ and $3/4$ positioning of the daughter nucleoids has been ubiquitously  observed in experiments \cite{wu2019cell}.

In what follows, we present a simple argument to explain the dependence of the length at which the nucleoid splits with respect to the underlying parameters, e.g.,  the mRNA synthesis rate. We take the nucleoid to be a region with homogeneous DNA-segment concentration which extends from $x=-L_{\textrm{nucl}}/2$ to $x=L_{\textrm{nucl}}/2$, with interfaces that are perfectly sharp. The mRNAs synthesized within the nucleoid will diffuse until they reach the nucleoid boundaries and, because it is energetically favorable, they will then automatically escape the nucleoid and not return. As a result, the steady-state concentration of mRNAs within the nucleoid can be modeled by the following diffusion equation with a uniform source term due to mRNA synthesis and absorbing boundary conditions that represent mRNA escaping from the nucleoid:
\begin{equation}
D_n\frac{\partial^2 \rho_{\rm tot}(x)}{\partial x^2}=\alpha c_{\smdna}, \qquad \rho_{\rm tot}\left(\pm \frac{L_{\textrm{nucl}}}{2}\right)=0,
\end{equation}
where $\rho_{\rm tot}(x)$ is the total mRNA concentration at position $x$, $D_n$ the mRNA diffusion constant (as defined in \textit{Model parameters}), and $\alpha$ the rate of synthesis of mRNA. The solution to the above equation is $\rho_{\rm tot}(x)=(L_{\textrm{nucl}}^2/2-x^2)\alpha c_{\smdna}/D_n$, whose maximum at $x=0$ is $L_{\textrm{nucl}}^2\alpha c_{\smdna}/(2D_n)$. 
We hypothesize that when the mRNA concentration at the center becomes larger than a given threshold, $\rho_{\rm tot}^*$,  spinodal decomposition takes place due to steric interactions between mRNAs and DNA, causing the nucleoid to split into two lobes. We thus expect $\rho_{\rm tot}^*$ to roughly correspond to the spinodal line of the phase diagram, but, given the out-of-equilibrium nature of the system due to, e.g. mRNA synthesis, it could differ from the equilibrium spinodal boundary.  Whatever value $\rho_{\rm tot}^*$ takes (provided its dependency on $\alpha$ is negligible), this simple model predicts a scaling for the critical length $L_{\textrm{nucl}}^*$ at which the nucleoid starts to divide  of the form $L_{\textrm{nucl}}^* \propto (\alpha c_{\smdna})^{-1/2}$, obtained from equating the maximum of the mRNA concentration profile to a fixed value  $\rho_{\rm tot}^*$. To test the prediction of this simple model, we numerically obtained the length at which the nucleoid divides for different values of $\alpha$, see the inset in Fig. \ref{fig1}D, and found a good agreement with the proposed scaling.

\subsection*{Single-chromosome filamentous growth}

So far we have analyzed the scaling of nucleoid size with cell size by assuming that the number of DNA segments is proportional to cell length. In this section we analyze another case of biological interest, namely, the case of a cell with a fixed amount of DNA and varying cell size. This scenario was recently analyzed in a dynamic imaging study of the \textit{E. coli} chromosome \cite{wu2019cell}, where the initiation of DNA replication and cell division were halted, yielding a single chromosome in a filamentously growing cell.  We model this case by fixing the number of DNA segments, but allowing the cell size to vary. In addition, the mRNA and ribosome number are no longer be proportional to cell length: based on the data in Ref. \cite{hanna20xx}, we assume that the total concentrations of mRNAs and ribosomes decrease linearly with cell length, approaching zero at $30\, \mu {\rm m}$---see \textit{SI Appendix}, Section  \ref{scaling_single_chromosome} for details.

Results  are shown in Fig. \ref{fig2}: our model again predicts a roughly linear scaling of the nucleoid size with respect to cell length, while the DNA segment concentration decreases with cell size. This indicates that the decrease in DNA-segment concentration with cell size is balanced  by the decrease of mRNA and ribosome concentrations, so as to keep nucleoid size a linear function of cell size. This can be seen clearly in Fig. \ref{fig2}A where the concentrations of all components of the TTM decrease as the cell size increases. While the model prediction for nucleoid versus cell length agrees reasonably well with experiments \cite{wu2019cell} for cell lengths smaller than $\sim 10 \, \rm \mu m$, there is a  discrepancy for larger cells, see \textit{Discussion}. 

\begin{figure}
\includegraphics[scale=1.45]{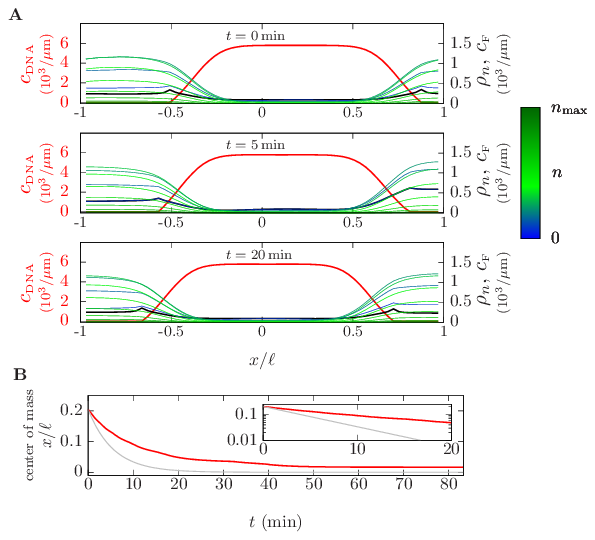}
\figCaption{ \label{fig5} Out-of-equilibrium processes center the nucleoid at midcell}{(\textit{A}) Concentration profiles obtained by initially shifting the steady-state profiles towards the right cell pole at $t=0$, and then integrating forward in time the reaction-diffusion \crefrange{RD_eqs_C}{reac_diff} in the presence of out-of-equilibrium processes to $t= 5 \, \rm min$ and $t =20\, \rm min$, for a cell with half-length $\ell =  1.8 \, \mu \rm m$, shown as in Fig. \ref{fig1}C. (\textit{B}) In red, location of the center of mass of the nucleoid along the long cell axis, as a function of time. In gray, the analytical lower bound obtained by neglecting nucleoid drag. Inset: The quantities depicted are the same as in \textit{B}, with the \textit{y}-axis is in logarithmic scale.
}
\end{figure}

\subsection*{Nucleoid centering}

As observed in Ref. \cite{wu2019cell}, a single bacterial nucleoid has a strong tendency to localize at midcell  for all cell sizes. Following the recent suggestion that the  central positioning of the nucleoid is regulated by an active process \cite{joyeux_microorganism}, we investigated whether the out-of-equilibrium process of mRNA production, diffusion, ribosome binding, and mRNA degradation can account for  nucleoid centering.

We consider the case of a nucleoid that, due to a fluctuation, is not initially at the center of the cell, and test whether the out-of-equilibrium effects in our model can push the nucleoid back to the cell center. To model this, we use the steady-state profiles obtained for filamentous growth, and shift  the concentration profiles towards the right cell pole. 
The resulting configuration has a nucleoid displaced from the center, and equal mRNA and ribosome concentrations on both sides of the nucleoid. This concentration profile is used as the initial condition for \crefrange{RD_eqs_C}{reac_diff}, which we integrate forward in time in the presence of the out-of-equilibrium terms. As shown in Fig. \ref{fig5} and \textit{Movie S2}, the nucleoid is centered at midcell after $\sim 30 \, \rm min$. 

The physical origin of this centering is mRNA synthesis in the nucleoid: The nascent mRNAs diffuse in the nucleoid until they reach one of its boundaries and then escape, with an equal flux to the left and right of the nucleoid. If the nucleoid is not centered, the accumulating mRNAs occupy a greater fraction of the available volume on one side of the nucleoid and thus create a higher osmotic pressure on that side. The resulting pressure difference ultimately drives the nucleoid back to the center of the cell.

The rate at which the nucleoid moves toward the cell center depends on  both the pressure difference due to mRNA accumulation, and on the effective viscous drag experienced by the nucleoid. We can establish a lower bound for the time it takes the nucleoid to center by assuming that the response of the nucleoid to an osmotic pressure difference is fast (low drag), such that the nucleoid is always located at a position where the osmotic pressure difference vanishes. Then, the centering process is only limited by the speed at which mRNAs accumulate on either side of the nucleoid, which sets the pressure differences. The kinetics obtained in this limit are shown in Fig. \ref{fig5}B, and they are given by an exponential relaxation with timescale $\beta^{-1}$, set by the rate of mRNA degradation (see  \textit{SI Appendix}, Section  \ref{estimates}). As shown in the Figure, the nucleoid centering obtained from the full model lags behind the lower bound, showing that there is a non-negligible contribution from drag on the nucleoid. Both the lower bound and the result from the full model show an exponential relaxation of the nucleoid position for early times in the centering process.

\subsection*{Nucleoid expansion due to halt of mRNA synthesis}
\begin{figure}
\includegraphics[scale=1.45]{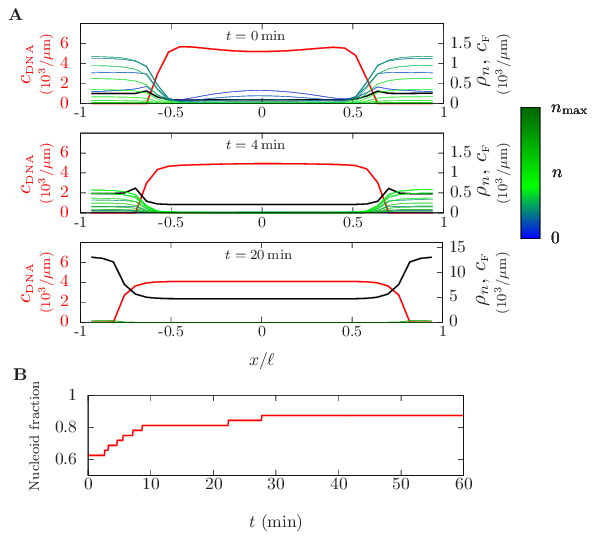}
\figCaption{ \label{fig4} The nucleoid expands in the absence of mRNAs synthesis}{(\textit{A}) Steady-state profile including mRNA synthesis ($t=0$) and profiles obtained by integrating forward in time from 
the steady-state profile at $t=0$ in the absence of mRNA synthesis ($t =4 \, \rm mins$ and $t =20 \, \rm mins$), for a cell with half length $\ell =  3.6 \, \mu \rm m$. The concentration profiles are shown as in Fig. \ref{fig1}C. 
(\textit{B}) Fraction of the cell volume occupied by the nucleoid, as a function of time (computed as the fraction of length along the axis of the cell with a DNA segment concentration $c_\smdna > 1000\, \, \mu\textrm{m}^{-1})$. The step-like shape of the graph is due to the spatial discretization in our numerical solutions (see \textit{SI Appendix}, Section  \ref{numerics}).
}
\end{figure}

It has been shown experimentally that when \textit{E. coli} transcription is halted, e.g. by treatment with rifampicin, the nucleoid expands \cite{bakshi2012superresolution,Cabrera2009ActiveTranscription}. A halt of mRNA synthesis depletes polysomes, and thus results in a lower osmotic pressure on the nucleoid. We tested this scenario with our model by using the out-of-equilibrium steady state shown   in Fig. \ref{fig1}C as the initial condition for  \crefrange{RD_eqs_C}{reac_diff}, switching off mRNA synthesis, and integrating forward in time. As shown in Fig. \ref{fig4}, the nucleoid expands and spreads over most of the intracellular space. The nucleoid does not take over the entire cell because there are pockets of free ribosomes at both cell poles, which prevent the DNA from occupying these spaces.

The nucleoid reaches its expanded steady state in $\sim 30 \, \rm min$, which is in good agreement with experimental data \cite{Cabrera2009ActiveTranscription}. However, the bulk of the expansion happens in the first $10 \, \rm min$---a timescale consistent with the half-life of mRNA ($5 \, \rm min$), whose degradation drives the expansion process.

\section*{Discussion}

In this study, we investigated the physical origins of the intracellular localization of DNA, messenger RNAs (mRNAs), and ribosomes in bacteria. This is a topic of general interest due to its far-reaching consequences, e.g.,  the spatial organization of transcription and translation \cite{gray2019nucleoid, Surovtsev2018SubcellularOrganization, Weng2019SpatialRNAP}, chromosome positioning   and segregation  \cite{joshi2011escherichia,wu2019cell}, and a wide range of cellular processes regulated by the nucleoid that excludes many  macromolecules from the volume which it occupies \citep{Coquel2013LocalizationProtein,Janissen2018DNACompaction}.

We developed a model for the spatial organization of the bacterial nucleoid based on steric interactions among DNA, mRNAs, and ribosomes. The model predicts the formation of a phase-separated nucleoid, whose size is in agreement with experimental measurements \cite{wu2019cell} for cells smaller than $10\, \mu\textrm{m}$ (Fig. \ref{fig1}). Beyond this cell length, our model is no longer accurate, for reasons that may include the lack of connectivity among modeled DNA segments, uncertainties in the concentration of crowders, and molecular components not considered in the model, such as nucleoid-associated proteins \cite{Dame2020} or topoisomerases that control DNA supercoiling \cite{Stuger2002}. The model also accounts for nucleoid expansion as a result of a halt in mRNA synthesis, demonstrating that the progressive degradation of crowders could be the physical cause of the expansion. Indeed, the timescales on which such expansion happens matches the one observed experimentally \cite{Cabrera2009ActiveTranscription}, and coincides with the timescales of mRNA turnover.

Our results underline the importance of out-of-equilibrium effects in the regulation of  nucleoid size and position. The nucleoid is known to localize at midcell \cite{wu2019cell}, and we demonstrate that the synthesis of mRNAs and their expulsion from the nucleoid caused by steric effects is sufficient to give rise to this positioning---see Fig. \ref{fig5}. In fact, a perturbation from the central position of the nucleoid induces an osmotic-pressure difference between the two cell poles, which pushes the nucleoid back to midcell. The timescale for this centering depends on both the time it takes to establish an osmotic-pressure difference, which is set by the mRNA turnover time, and the drag experienced by the nucleoid.  This drag may be underestimated in our model, because we do not include effects that could slow down  nucleoid centering, e.g., the transient attachment of the nucleoid to the membrane by proteins that are simultaneously being transcribed, translated, and inserted in the membrane, also known as transertion \cite{Gorle2017}. 
Furthermore, our model shows that out-of-equilibrium effects are responsible for the ubiquitous nucleoid splitting and localization at $1/4$ and $3/4$ positions along the long cell axis. Thus, our analysis shows that the synthesis of mRNAs within the nucleoid, without additional active processes, is a robust mechanism to make the daughter nucleoids localize at $1/4$ and $3/4$ positions, as observed experimentally \cite{wu2019cell}. 

Our study implies that steric interactions make the bacterial cytoplasm a poor solvent for the chromosome, as recently indicated by experiments \cite{Xiang2020SolventQuality}. However, steric interactions may not be the only contribution to the poor-solvent quality of the cytoplasm. Other types of intermolecular interactions \cite{Odijk1998OsmoticCompaction} or the effect of nucleoid-associated proteins \cite{Dame2020} could also affect the solvent quality of the cytoplasm and the organization of the nucleoid in the cell. For future studies, both theoretical and experimental, research into these other regulators of the nucleoid size could yield a more complete picture of its organization, and improve the accuracy of the results presented here. 

Finally, we observe that Turing patterns \cite{Turing1952patterns} display out-of-equilibrium patterning features that could seem similar to the ones produced by our model, see \textit{SI Appendix} Fig. \ref{fig_lobes}. These out-of-equilibrium patterns have been used to investigate many biological features on a cellular scale, such as the positioning of protein clusters in \textit{E. coli} \cite{murray2017self}. However, unlike Turing patterns, our model predicts phase separation exclusively due to steric interactions and in the absence of out-of-equilibrium effects, see Fig. \ref{fig3}.  While the patterns produced by our model could be related to other out-of-equilibrium phase-separation models \cite{Li2020noneq} such as models of growing droplets \cite{Zwicker2017division}, our model provides a  conceptually simpler framework to produce these patterns. In fact, unlike a model of physically growing droplets,  our analysis involves a conserved order parameter---the total number of DNA segments---and the effect of out-of-equilibrium terms---mRNA production and degradation---is limited to nucleoid reshaping and division. Despite its simplicity, our model produces a number of experimentally observed patterning effects, such as  nucleoid centering at midcell,  and splitting and positioning of sister lobes during cell division. In addition, the patterning of our model is not limited to nucleoid splitting into two sister lobes, because our model predicts that the nucleoid can split into more than two lobes, whose size is given by a characteristic length and whose positions are tightly controlled, see \textit{SI Appendix}, Fig.  \ref{fig_lobes}. Similar patterns for ordered nucleoid positioning have been found in long filamentously growing \textit{E. coli} cells \cite{Wehrens2018}, albeit with a shorter characteristic length.

Given its generality, our analysis is not restricted to the nucleoid of prokaryotic cells \cite{Pappu2020Condensates}. Indeed,  division of certain phase-separated condensates has been experimentally related to out-of-equilibrium processes, as is the case for the ParABS partition system, which creates phase-separated condensates of DNA and ParB, around \textit{par}S sites, whose  division is controlled by the activity of ParB's ATPase activity on ParA \cite{Guilhas2020ATPDriven}. 
The activity-driven nucleoid division described in our model may thus constitute a general strategy employed by cells to control the structure of membraneless compartments.

\begin{acknowledgments}

We thank J. Prost, A. Sclocchi, P. Sens, H. Salman, and J. Wagner for valuable conversations and suggestions. This study was supported in part by Agence nationale de la recherche  (ANR), grant ANR-17-CE11-0004, and the National Science Foundation, through the Center for the Physics of Biological Function (PHY-1734030). 

\end{acknowledgments} % Display the acknowledgments section

\onecolumngrid
\appendix

\section{Binary mixture of hard spheres} \label{toy_model}

In this Appendix we present a toy model to illustrate the concepts of the full \textit{E. coli} reaction-diffusion model, \crefrange{RD_eqs_C}{reac_diff}. The model is composed of $N_\textrm{A}$ and $N_\textrm{B}$ particles of species ${\rm A}$ and $\rm B$, which are hard spheres with radii $R_{\rm A}$, $R_{\rm B}$ and diffusion coefficients  $D_{\rm A}$, $D_{\rm B}$, respectively, confined in a volume $\mathcal{V}$.  We assume that the system has cylindrical symmetry, and thus $\mathcal{V}$  is effectively one dimensional, so as to apply our findings to \crefrange{RD_eqs_C}{reac_diff}. 

%Denoting by $c_i(x,t)$ the concentration at point $x$ and time $t$ of species $i$, in the absence of excluded-volume effects the particle current would be proportional to $-\partial_x c_i(x,t)$. 
%There is a hard-core  interaction potential between any particle pair, which vanishes when $r_{ij} \geq R_i+R_j$, where $r_{ij}$ is the distance between the centres of particle $i$ and particle $j$, and is infinitely large otherwise. 

\subsection{Free energy}

We use the virial expansion, first developed by Onnes \cite{kamerlingh1902knaw} over a century ago, to compute the free energy of a gas of interacting particles. Here we limit ourselves to state the result for hard-sphere potentials to second and third order and refer the interested reader to classical statistical physics textbooks (for instance, \cite{huang1987statistical,Pathria1996statistical}) for a complete explanation of the procedure.

The hard-sphere potential $V_{ij}$ between the $i$th and $j$th particle is zero if the distance between particles is larger than the sum of their radii and infinity otherwise. Then, the free energy of the system is
\begin{align}
F=&-k_{\rm B} T \log Z \nn \\
 =& \, k_{\rm B} T   \sum_{a=\rm{A,B}} N_a \log \frac{N_a \Lambda_a^3}{\mathcal{V}}  \nn \\ 
& - k_{\rm B}T\log \Bigg[ 1 -  \sum_{a=\textrm{A,B}} \frac{N_a(N_a-1)}{2\mathcal{V}} B^{(2)}_{aa}-\frac{N_{\textrm{A}}N_{\textrm{B}}}{\mathcal{V}}   B^{(2)}_{\textrm{AB}}-\sum_{a=\textrm{A,B}}  \frac{N_a(N_a-1)(N_a-2)}{6\mathcal{V}^2}B^{(3)}_{aaa}\\ \nn
& - \sum_{a \neq b}\frac{N_a(N_a-1)N_b}{2\mathcal{V}^2}B^{(3)}_{aab} \Bigg] ,
\end{align}
where $k_{\rm B}$ is the Boltzmann constant, and $\Lambda_a$ is the thermal de Broglie wavelength of species $a$, and the $i$th virial coefficients $B_{ab}^{(i)}$ are given by:
\begin{align}
\label{virial_sph}
B^{(2)}_{ab}&=\frac{4 \pi }{3} ( R_{ a} + R_{ b})^3, \\ B^{(3)}_{abc}&=\frac{16 \pi^2}{9}  \left[ R_b^3 R_c^3 + 3 R_a R_b^2 R_c^2 (R_b + R_c) + R_a^3 (R_b + R_c)^3 + 3 R_a^2 R_b R_c (R_b^2 + 3 R_b R_c + R_c^2)\right].
\end{align}

 Assuming that the  volume is large,  we can approximate the free energy as
\begin{align}\nn
F \simeq & \, k_{\rm B} T  \Bigg[  \sum_a N_a \log \frac{N_a \Lambda_a^3}{\mathcal{V}}  +\sum_{a=\textrm{A,B}} \frac{N_a(N_a-1)}{2\mathcal{V}} B^{(2)}_{aa}+\frac{N_{\textrm{A}}N_{\textrm{B}}}{\mathcal{V}}   B^{(2)}_{\textrm{AB}} \\ 
& +\sum_{a=\textrm{A,B}}  \frac{N_a(N_a-1)(N_a-2)}{6\mathcal{V}^2}B^{(3)}_{aaa}+ \sum_{a \neq b}\frac{N_a(N_a-1)N_b}{2\mathcal{V}^2}B^{(3)}_{aab} \Bigg],
\end{align}
where we have used the expansion of the logarithm, which is accurate for small concentrations.

Now we consider an infinitesimal distance $d x$, in which there is an infinitesimal number of molecules of each species $d N_a$, with one-dimensional concentration 
$c_a(x)=dN_a/dx$. The volume of each of these infinitesimal slices of the system is $dV = \sigma \,dx$, where $\sigma$ is the cross section of the system. Then,
\begin{align}
\label{unif}
F_0=& \int_{-\ell}^\ell d F_0 \nn \\
\simeq&\, \int_{-\ell}^\ell k_{\rm B} T d x   \Bigg\{ \sum_{a=\textrm{A,B}} c_a(x) \left[ \log c_a(x) + \log \frac{\Lambda_a^3}{\sigma}\right] +  \sum_{a=\textrm{A,B}} \left[ B^{(2)}_{aa} \frac{[c_a(x)]^2}{2 \sigma} +  B^{(3)}_{aaa}\frac{[c_a(x)]^3}{6\sigma^2} \right]\nn\\ 
& +    B^{(2)}_{\textrm{AB}} \frac{c_{\textrm{A}}(x)c_{\textrm{B}}(x)}{\sigma}+\sum_{a \neq b}\frac{[c_a(x)]^2c_b(x)}{2\sigma^2}B^{(3)}_{aab} \Bigg\},
\end{align}
where we have approximated $N_a-1$ with $N_a$.

The quantity $dF_0/dx$ in  \cref{unif} is the free-energy density of the system assuming the concentrations are uniform along the $x$ axis, where this condition is denoted by the subscript `$0$'. If we assume the true, local free energy density of the system  
$f=dF/dx$, to be a function of the uniform free-energy density and the derivatives of the concentration, i.e., $f=f(f_0, \nabla c_a, \nabla^2 c_a...)$, then we can expand it around $f_0$, considering the concentration and its derivatives as independent variables, as follows \cite{cahn1958free}:
\begin{equation}
\label{Cahn_eq}
f= \, f_0 + \sum_a \gamma_a \frac{d^2c_a}{dx^2} + \frac{k_\textrm{B}T}{2}\sum_{a,b} \frac{\kappa_{ab}}{\sigma} \frac{dc_a}{dx}\frac{dc_b}{dx},
\end{equation}
where $\kappa_{ab}$ are the Cahn-Hilliard coefficients, which account for spatial inhomogeneities in the concentrations. 
The second term in the right-hand side of \cref{Cahn_eq} does not contribute to the total free energy of the system: indeed, when spatially integrated, this term vanishes because of the Neumann boundary conditions. Then, the total free energy is
\begin{equation}\label{eq_F}
F = \int_{-\ell}^\ell d F = \int_{-\ell}^\ell d F_0 + \frac{k_{\rm B}T}{2}\int_{-\ell}^\ell \sum_{a,b} \kappa_{ab} \frac{dc_a}{dx}\frac{dc_b}{dx} d x
\end{equation}
and the chemical potential for, e.g., species ${\rm A}$ reads
\begin{align}\nn 
\mu_{\textrm{A}}(x)=&  \,\frac{\delta F}{\delta c_{\textrm{A}}(x)} \\\nn
 =&  \, k_{\rm B}T \left[ 1 + \log c_{\textrm{A}}(x) + \frac{c_{\textrm{A}}(x)}{\sigma} B^{(2)}_{\textrm{AA}} +\frac{c_{\textrm{B}}(x)}{\sigma} B^{(2)}_{\textrm{AB}} - \sum_{a=\textrm{A,B}} \kappa_{\textrm{A}a}\frac{d^2c_a(x)}{dx^2}  \right] \\ \label{eq_mu}
&+k_{\rm B}T\left[\frac{c_{\textrm{B}}(x)c_{\textrm{A}}(x)}{\sigma^2} B^{(3)}_{\textrm{AAB}}+\frac{c_{\textrm{A}}(x)c_{\textrm{A}}(x)}{2 \sigma^2} B^{(3)}_{\textrm{AAA}}+\frac{c_{\textrm{B}}(x)c_{\textrm{B}}(x)}{2\sigma^2} B^{(3)}_{\textrm{BBA}}\right], 
\end{align}
where $\kappa_{aa}$ has absorbed a factor of 2 arising from the derivative of the second term in \cref{eq_F} when $a=b$. For future convenience, we define $\nu_a$ as the non-ideal contribution to the chemical potential of each species,  e.g., species ${\rm A}$ :
\begin{equation}
\label{excl_vol}
\nu_{\textrm{A}}(x)= \frac{B^{(2)}_{\textrm{AA}}}{\sigma}c_{\textrm{A}}(x) +\frac{ B^{(2)}_{\textrm{AB}}}{\sigma} c_{\textrm{B}}(x)+ \frac{B^{(3)}_{\textrm{AAB}}}{\sigma^2}c_{\textrm{B}}(x)c_{\textrm{A}}(x)+\frac{B^{(3)}_{\textrm{AAA}} }{2\sigma^2}[c_{\textrm{A}}(x)]^2 +\frac{B^{(3)}_{\textrm{BBA}}}{2 \sigma^2}[c_{\textrm{B}}(x)]^2 - \sum_{a=\textrm{A,B}} \kappa_{\textrm{A}a}\frac{d^2c_a(x)}{dx^2}.
\end{equation}

\subsection{Currents}
 
We can now work out the currents by considering Fick's law of diffusion, namely,
\begin{equation}
\label{Ficks}
J_a(x)=-\frac{D_a(x)}{k_\textrm{B} T} c_a(x) \partial_x \mu_a,
\end{equation} 
where $D_a(x)$ is the diffusion constant and is given by Einstein-Smoluchowski-Sutherland relation \cite{dill2012molecular}
\begin{equation}
D_a(x)=\zeta_a(x) k_\textrm{B} T,
\end{equation}
and $\zeta_a(x)$ is the spatially dependent mobility of species $a$. 

%Guided by the general idea that diffusion of one particle is hindered by its excluded-volume interactions with the ensemble of the other particles, we will make the following ansatz for the diffusivity:
%\begin{equation}
%\label{diff}
%D_a(x)=D_a^\ast \frac{v_a(x)}{v^\ast_a}= D_a^\ast e^{-\nu_a (x)+\nu^\ast_a},
%\end{equation} 
%where $\nu_a(x)$ is the component of the chemical potential (\ref{eq_mu}) which accounts for excluded-volume interactions only, i.e., 
%\begin{equation}
%\label{excl_vol}
%\nu_{\textrm{A}}(x)= \frac{B^{(2)}_{\textrm{AA}}}{\sigma}c_{\textrm{A}}(x) +\frac{ B^{(2)}_{\textrm{AB}}}{\sigma} c_{\textrm{B}}(x)+ \frac{B^{(3)}_{\textrm{AAB}}}{\sigma^2}c_{\textrm{B}}(x)c_{\textrm{A}}(x)+\frac{B^{(3)}_{\textrm{AAA}} }{2\sigma^2}[c_{\textrm{A}}(x)]^2 +\frac{B^{(3)}_{\textrm{BBA}}}{2 \sigma^2}[c_{\textrm{B}}(x)]^2 - \sum_{a=\textrm{A,B}} \kappa_{\textrm{A}a}\frac{d^2c_a(x)}{dx^2}  ,
%\end{equation}
%and $\ast$ refers to the uniform configuration, where all concentration profiles are spatially uniform. The term $v^\ast_a$ in \cref{excl_vol} has been introduced so as to directly relate $D_a^\ast$ to measurable quantities. In fact, when evaluating both sides of \cref{excl_vol} at the uniform configuration, the constant $D_a^\ast$ coincides with the diffusivity $D_a(x)$ at the uniform configuration, which can be estimated experimentally---see section \ref{parameters}. 

By substituting  the expression \eqref{eq_mu} for the chemical potential in \cref{Ficks}, we obtain the current for species $\rm A$:
\begin{align}\nn
J_{\textrm{A}}(x) =& - D_{\textrm{A}} (x) \Bigg\lbrace  \partial_x c_{\textrm{A}}(x) + c_{\textrm{A}}(x) \sum_{a=\textrm{A,B}} \left[\frac{B^{(2)}_{\textrm{A}a}}{\sigma} \partial_x c_{\textrm{a}}(x) - \kappa_{Aa}\partial_x^3 c_{\textrm{a}}(x) \right] + \frac{B^{(3)}_{\textrm{AAA}}}{\sigma^2} [c_{\textrm{A}}(x)]^2\partial_x c_{\textrm{A}}(x) \\ \label{excl_vol_curr}
 &+ \frac{B^{(3)}_{\textrm{AAB}}}{\sigma^2}\left([c_{\textrm{A}}(x)]^2 \partial_x c_{\textrm{B}}(x)+c_{\textrm{A}}(x)c_{\textrm{B}}(x)  \partial_x c_{\textrm{A}}(x)\right)+ \frac{B^{(3)}_{\textrm{BBA}}}{\sigma^2}c_{\textrm{A}}(x)c_{\textrm{B}}(x)  \partial_x c_{\textrm{B}}(x)\Bigg \rbrace.
\end{align}
Proceeding along the same lines, we obtain the current for species $\rm B$. By using these currents, we will work out the diffusion equations of the system in the following section.

\subsection{Free-energy minimum and equilibrium steady state}\label{steady}

The diffusion equations for the binary mixture of hard spheres discussed in section \ref{toy_model} read
\be\label{eqAB}\left\{
\begin{array}{ccc}
\partial_t c_{\rm A}(x,t) &= & - \partial_x J_{\rm A}(x,t),\\
\partial_t c_{\rm B}(x,t) & = & - \partial_x J_{\rm B}(x,t).
\end{array}
\right.
\ee

We will consider the steady state of \cref{eqAB} combined with no-flux boundary conditions, and with a constraint fixing the total number of particles to a given value, $N_a$, for each species:
\begin{align}\label{eq_1}
- \partial_x J_{a}(x,t) =  0, \,J_{a}(\pm \ell,t) = 0,\;\;& a = {\rm A,B},\\ 
\int_{-\ell}^\ell dx \, c_{a}(x, t)  =  N_a, \;\; &a = {\rm A,B}. \label{eq2}
\end{align}
We will show that Eqs.  \cref{eq_1,eq2} are equivalent to finding the minimum of the free energy with the constraint \eqref{eq2}, i.e.,
\begin{align}\label{minf} 
&\min_{c_{\rm A}, c_{\rm B}} F   \\\label{constr}
&\textrm{subject to Eq. } \eqref{eq2},
\end{align} 
\noindent
where $F$ is now considered to be a functional of the concentration profiles $c_a(x)$. 

The Lagrange function of the minimization problem given by  \cref{minf,constr} reads
\be
{\cal L} = F - \sum_a \lambda_a \left[  \int_{-\ell}^\ell dx \, c_{a}(x)  - N_a \right],
\ee
where $\lambda_a$ are the Lagrange multipliers. First, the stationarity condition of $\cal L$ with respect to $c_a(x)$ is given by
\begin{align}\nn
0 =&\, \frac{\delta {\cal L}}{\delta c_a(x)} \\ \nn
=&\, \frac{\delta F}{\delta c_a(x)} - \lambda_a \\\label{stat_La}
 = & \, \mu_a(x) - \lambda_a,
\end{align}
where in the third line we used the definition \eqref{eq_mu} of the chemical potential. By taking the derivative of \cref{stat_La} with respect to $x$ and using  \cref{Ficks}, we obtain $J_{\rm A}(x) = 0$, which is equivalent to \cref{eq_1}. Second, the stationarity condition of $\cal L$ with respect to $\lambda_a$ yields \cref{eq2}. As a result, the minimization problem \eqref{minf}, \eqref{constr} is equivalent to conditions \eqref{eq_1} and \eqref{eq2}. 

In what follows, we will seek the steady state \eqref{eq_1} with the conditions \eqref{eq2} on particle numbers, by solving the minimization problem  \eqref{minf},   \eqref{constr}. In fact, while a direct solution of  \cref{eq_1,eq2} may yield concentration profiles corresponding to locally stable free-energy minima,  the minimization method allows us to select the true, physical minimum of the free energy. 

\section{Currents for the \textit{E. coli} transcriptional-translational machinery} \label{currents_model}

We now apply the ideas discussed in section \ref{toy_model} to the reaction-diffusion model for the nucleoid, \crefrange{RD_eqs_C}{reac_diff}. 

In particular, we will apply the procedure discussed for the binary mixture of hard spheres to obtain an expression for the current along the lines of \cref{excl_vol_curr}. However, there is an important difference that need to be taken into account: Not all the particles are spheres, because the DNA segments are considered to be cylinders of length $L$ and radius $\rho$.

\begin{figure}
\begin{center}
\includegraphics[scale=0.5]{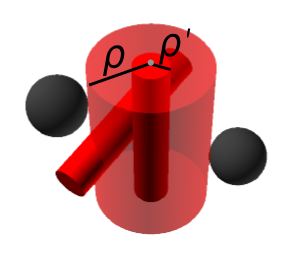}
\figCaption{Steric interactions of DNA segments}{\label{steric_cylinder} 
Hard-core interactions between DNA segments (cylinders, red) and ribosomes or polysomes (spheres, black). DNA segments interact with each other through a cylinder of radius $\rho'$ but interact with other polysomes and ribosomes with a radius $\rho > \rho'$, satisfying the relationship \eqref{rhorhop}.
}
\end{center}
\end{figure}

 In addition, as stated in the main text (see \textit{Results}), a DNA segment does not interact with another DNA segment in the same way in which it would interact with a polysome or ribosome. Therefore, we consider that a DNA segment interacts with other DNA segments as a cylinder of radius $\rho'< \rho$. We base the value of $\rho'$ on the hard-sphere model for DNA used for numerical simulations  \cite{mondal2011entropy}, where each plectoneme segment is represented as a sequence of four bond beads and two node beads, and all beads have radius $\rho$.  In the simulations, whenever a DNA segment collides with a particle which is not a DNA segment, none of the beads are allowed to overlap with the particle. On the other hand, whenever two DNA segments collide, node beads cannot overlap with each other, but bond beads can, according to the  picture above. Given that the two node beads are located at the vertices which connect segments, each node bead contributes half of its volume to each plectoneme segment. The volume that a DNA segment excludes to other DNA segments, which we denote by $\pi \rho'^2 L$, is thus the volume of one node bead, i.e., one fifth of the volume $\pi \rho^2 L$ that the segment excludes to particles other than DNA. As a result, we obtain the relation
\be\label{rhorhop}
\rho' = \rho/\sqrt{5}
\ee 
 between $\rho$ and $\rho'$. See Fig. \ref{steric_cylinder} for a sketch of this interactions.
 
\subsection{Virial coefficients for DNA segments} \label{DNA_virial}
 
Given the shapes of the particles discussed in the main text, section \textit{Model parameters}, the functional form of the contributions of steric effects to the virial expansion 
$B^{(i)}_{ab}$ for ribosomes and mRNA remains the same as in section \ref{toy_model}, but the interaction of other species with DNA, and of DNA with itself, is different. The second virial coefficient for two cylinders is \cite{herold2017virial}
\be
B^{(2)}_{\smdna \, \smdna}=2 \pi \rho' \left[ L \rho' + \frac{1}{2} (L+ \rho') (L+\pi \rho') \right], 
\ee
while the coefficient for one cylinder and a sphere with radius $R_n$ is
\be
B^{(2)}_{{\smdna} \, n}= L \pi (R_n+\rho)^2 + 2\pi R_n \left( \rho^2 + \frac{\pi \rho R_n}{2} + \frac{2}{3}R_n^2 \right),
\ee
where the subscript DNA stands for a DNA cylinder, and $n$ for a sphere of radius $R_n$. 

Before we present the expressions for the third virial coefficients for cylinders, let us define 
\begin{equation}
u_{ij}=\mathbb{I}(\textbf{q}_i \cap \textbf{q}_j),
\end{equation}
where  $\textbf{q}_i$ denotes the degrees of freedom which specify the position and orientation of a particle, $\textbf{q}_i \cap \textbf{q}_j$ stands for the condition that  particles $i$ and $j$ overlap, i.e., their hard-core potential is nonzero, and the indicator function $\mathbb I$  is equal to one if the condition in its argument is satisfied, and zero otherwise. The third virial coefficients for interactions that involve cylinders are given by the following integral expressions:
\begin{align}
B^{(3)}_{\smdnas \,n\, m}=& \, \int d \textbf{r}_{n\smdnas }d \textbf{r}_{m\smdnas } u^z_{\smdnas \, m}u^z_{\smdnas \, n}u_{nm} ,  \\
B^{(3)}_{\smdnas \, \smdnas'\, n}=&\, \frac{1}{8\pi^2} \int d \textbf{r}_{\smdnas \, \smdnas} d \textbf{r}_{\smdnas \, n} d \textbf{r}_{\smdnas' \, n}d (\cos \theta_\smdnas' )d (\cos \theta_{\smdnas'})d (\cos \theta_n)d \phi_{\smdnas}d \phi_{\smdnas'}d \phi_nu^z_{\smdnas \, \smdnas'}u^z_{\smdnas \, n}u_{\smdnas' \, n},  \\
B^{(3)}_{{\smdnas} {\smdnas'} {\smdnas''}}=&\, \frac{1}{8\pi^2} \int d \textbf{r}_{\smdnas \smdnas'}d \textbf{r}_{\smdnas' \smdnas''}d \textbf{r}_{\smdnas \smdnas''}d (\cos \theta_\smdnas)d (\cos \theta_{\smdnas'})d (\cos \theta_{\smdnas''})d \phi_\smdnas d \phi_{\smdnas'}d \phi_{\smdnas''}u^z_{\smdnas \smdnas'}u^z_{\smdnas \smdnas''}u_{\smdnas' \smdnas''} ,
\end{align}
where the subscripts label different cylinders, D stands for DNA,  $n$ and $m$ label the spheres, and vectors $\textbf{r}_{ij}$ denote the relative position  between the centers of mass of particles $i$ and $j$. In addition, the superscript $z$ means that the axis of cylinder D is parallel to the $z$ axis, so as to leverage spherical symmetry, and  $\theta$, $\phi$ are the polar and azimuthal angles, respectively. While some simplifications of those integrals are possible, there is no known analytical form for these virial coefficients  \cite{straley1973third}.

Then, the total free energy of the system is:
\begin{align}
\label{F_tot}
\frac{F}{k_{\rm{B}} T} = & \int_{-\ell}^\ell  dx \Bigg[ \frac{F_0}{k_{\rm{B}}T}+   \frac{B^{(2)}_{\smdna,\smdna}}{2\sigma} [c_{\smdna}(x)]^2+  \frac{B^{(3)}_{\smdna,\smdna,\smdna}}{6\sigma^2}[c_{\smdna}(x)]^3  +    \sum_{a=\textrm{F},n} \frac{B^{(2)}_{\smdna,a}}{\sigma} c_{\smdna}(x)c_{a}(x)+ \nn \\ & \sum_{\substack{a,b=\textrm{DNA,F},n \\ a \neq b}}\frac{B^{(3)}_{\smdna ,a,b} }{2\sigma^2} c_a(x)c_b(x)c_{\rm{DNA}}(x)\Bigg]  + \frac{1}{2}\int_{-\ell}^\ell  d x \sum_{a,b=\smdna,\smf,n} 
\kappa_{ab} \frac{dc_a}{dx}\frac{dc_b}{dx},
\end{align}
where the sums run over the chemical species denoted by F, DNA and all polysome species, which we denote by  `$n$' in the sums. Note that $F_0$ has the same structure as  \cref{unif} in section \ref{toy_model}: its form does not change because it involves only spherical-particle species. The difference between  $F_0$ in \cref{F_tot} and $F_0$ in \cref{unif} is in the summation indices, which now span over all polysome species and free ribosomes. 
As stated in the main text, the free energy of the system would include an entropic term related to the DNA to the free energy of the system, but, since this term is small compared with the steric interactions, we neglect it for simplicity (a more detailed explanation is given in section \ref{DNA_entropy}).
The steric interactions have been written explicitly for the DNA segments, and the Cahn-Hilliard terms are analogous to those of section \ref{toy_model}, except for the numerical values of the coefficients $\kappa_{ab}$, which depend on the particle geometry. 

In principle the Cahn-Hilliard coefficients $\kappa_{ab}$ in \cref{F_tot} can be computed by leveraging hard-core interactions as shown in Ref. \cite{Ilker2020phaseseparation}. However, for our purposes, it is enough to observe that the Cahn-Hilliard terms reflect the cost of concentration gradients related to species $a$ and $b$. It follows that  $\kappa_{ab}$ is an intrinsic feature of the particles of species $a $ and $b$: Because  $\kappa_{ab}$ has the dimension of the cube of a length, it must be proportional to a product of the linear sizes of the particles of species $a$ and $b$. In addition, the relation of the Cahn-Hilliard coefficients  to  differentials of concentrations over infinitesimal length scales indicates that $\kappa_{ab}$ is physically related to short rather than long length scales. We thus assume that $\kappa_{ab}$ equals the minimum between the volume of species $a$ and that of species $b$.

\subsection{ DNA free energy } \label{DNA_entropy}

The free energy of the DNA chain consists of two terms: An energetic and an entropic part. In our model of  DNA composed by disconnected cylindrical segments, the energetic part is given by the interactions between DNA segments, encoded by the virial-expansion terms, see  Section \ref{DNA_virial}. In this subsection we argue that the entropic part, can be neglected as it is smaller than the virial terms. The main contribution to the entropic part of the free energy can be captured by considering the free-energy cost of confining an ideal polymer. This entropic cost is given by \cite{Edwards1969confinement, degennes1979scaling}:
\begin{equation}
\label{entropy_dna}
S \simeq -k_\textrm{B} \frac{N_{\smdna} L^2}{D^2},
\end{equation}
where $D$ is the typical lengthscale on which the polymer is confined. In the case of the nucleoid, $D \sim 1\, \mu \textrm{m}$ and $N_\smdna=6\times 10^3$.

By inserting the numerical values of the parameters we obtain an entropic contribution of the order of $10^3 k_\textrm{B}T$ and a contribution from the virial terms of approximately 
\begin{equation}
k_\textrm{B}T \frac{N_{\smdna}^2  B_{\smdna,\smdna}^{(2)}}{2V_n} \sim 10^4 k_\textrm{B}T,
\end{equation}
 where $V_n \sim D^3$ is the typical volume of a nucleoid. Therefore, there is a difference of one order of magnitude between the entropic and the virial term and, for the sake of simplicity, in this work we neglect the entropic contribution to the DNA-segment current.

\subsection{Auxiliary entropy}

\begin{figure}
\begin{center}
\includegraphics[scale=1.5]{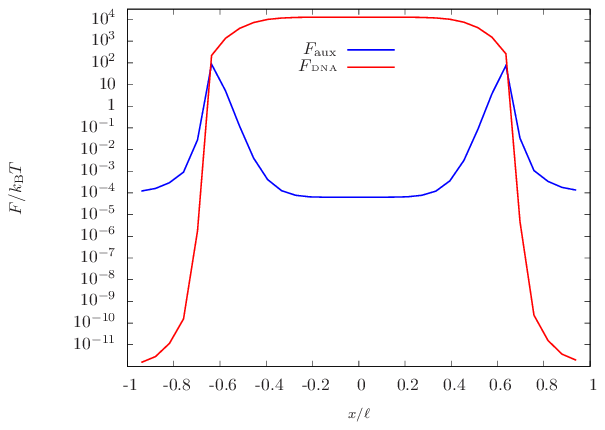}
\figCaption{Real and auxiliary free energies}{\label{entropies} 
Real and auxiliary free energies for a cell of size $3.6\,\mu {\rm m}$ at the out-of-equilibrium steady state. The real free energies only takes into account the self-interaction terms of DNA cylinders, as the mean-field entropic term of the DNA chain is neglected, as explained in section \ref{DNA_entropy}.
}
\end{center}
\end{figure}

In our analysis, we will first determine the steady state of the system in the absence of reaction and out-of-equilibrium terms, by minimizing the total free energy \eqref{F_tot}. These steady-state profiles will then be used as initial conditions to integrate forward in time the reaction-diffusion \crefrange{RD_eqs_C}{reac_diff}, which include both reaction and out-of-equilibrium terms. At the free-energy minimum, the DNA concentration is nonzero in the nucleoid, while it vanishes outside the nucleoid. Given that these equilibrium profiles are entered as initial conditions  in \crefrange{RD_eqs_C}{reac_diff}, the vanishing concentration above causes numerical instabilities  when these equations are numerically integrated forward in time, and can lead to negative concentrations in the out-of-equilibrium steady state \cite{shampine2005non}. To overcome this issue, we included a small, additional entropic term in the free energy \eqref{F_tot}:
\begin{equation}
\label{aux_terms}
F_{\rm aux} =K_{\textrm{aux}} \; k_{\rm B} T   \int_{-\ell}^\ell dx \, e^{-D_{\textrm{aux}}c_{\smdna}(x)/\langle c_{\smdna}(x)\rangle }\frac{c_{\smdna}(x)}{N_{\rm DNA}} \log 
\left[ 2 \ell  c_{\smdna}(x) \right],
\end{equation}
where 
$\langle c_{\smdna}(x) \rangle = N_{\rm DNA}/(2 \ell)$ is the average DNA concentration across the cell, and $K_{\textrm{aux}}$ and $D_{\textrm{aux}}$ are constants that we set to 
$0.2\,N_{\rm DNA}$ and $10$, respectively. The exponential term in \cref{aux_terms} is such that, if the coordinate $x$ lies in the nucleoid bulk, then $c_{\smdna}(x)$ is of the same order of magnitude as $\langle c_{\smdna}(x) \rangle$ and thus, by choosing $D_{\rm aux}$ sufficiently large, the contribution to $F_{\rm aux}$ vanishes. On the other hand, the exponential is approximately equal to one outside the nucleoid, where $c_{\smdna}(x) \ll \langle c_{\smdna}(x) \rangle$,  and \cref{aux_terms} reproduces a term that resembles the  standard entropy of mixing
\be\label{eq_entr}
\frac{c_{\smdna}(x)}{N_{\rm DNA}}  \log \left[ 2 \ell   c_{\smdna}(x) \right]
\ee 
of a polymer in the mean-field approximation, which dominates the integral in $F_{\rm aux}$. In what follows, we will add $F_{\rm aux}$ to the total free energy \cref{F_tot}, by setting
\be\label{F_tot_aux}
F \rightarrow F + F_{\rm aux}.
\ee
As a result, the minimization of \cref{F_tot} tends to also maximize the entropy \eqref{eq_entr}, thus spreading out a fraction of DNA segments outside the nucleoid bulk. This procedure alters only slightly the free-energy minimum: As shown in Fig. \ref{entropies}, the auxiliary free energy does not vanish outside of the nucleoid, but it is orders of magnitude smaller than the original free energy within the nucleoid.  Notwithstanding this, such a small free energy  prevents numerical instabilities in the integration of the reaction-diffusion equations. 

\subsection{Currents} 

Proceeding along the lines of section \ref{toy_model}, we obtain the DNA current from \cref{F_tot_aux}:
\begin{equation}
\label{curr_DNA}
J_{\smdna}(x) = -c_{\smdna}(x) D_{\smdna}(x) \left[ \partial_x \nu_{\smdna}(x) + \partial_x \mu_{\rm aux}(x)\right],
\end{equation}
where $\nu_{\smdna}$ is the excluded-volume term analogous to that in  \cref{excl_vol}. The term $\mu_{\rm aux}$ is the derivative of the auxiliary free energy \eqref{aux_terms} with respect to the DNA concentration, $c_{\smdna}(x)$, and its contribution to the current reads

\begin{align}
\partial_x \mu_{\rm aux}(x) =&\,K_{\rm aux} e^{-D_{\textrm{aux}}c_{\smdna}(x)/\langle c_{\smdna}(x)\rangle} \partial_x c_{\smdna}(x) \Bigg\{ \frac{1}{c_{\smdna} (x) } \left(1-D_{\textrm{aux}}\frac{c_{\smdna}(x)}{\langle c_{\smdna}(x)\rangle}\right) +D_{\textrm{aux}}\log [2 \ell \sigma c_{\smdna} (x) ]\nonumber \\ 
&- \frac{D_{\textrm{aux}}}{\langle c_{\smdna}(x)\rangle} \left[ 1 + \log [2 \ell \sigma c_{\smdna} (x)] \left(1-D_{\textrm{aux}}\frac{c_{\smdna}(x)}{\langle c_{\smdna}(x)\rangle}\right)\right] \Bigg\}.
\end{align}

Combining the results from section \ref{toy_model}, \cref{excl_vol_curr}, with those derived in this section, \cref{F_tot,curr_DNA}, we obtain the currents in \crefrange{RD_eqs_C}{reac_diff}. Making use of \cref{excl_vol_curr,curr_DNA} and of the virial coefficients previously derived, the currents for DNA, ribosomes and polysomes are fully defined.

We note that in the derivation of these currents we have incorporated a general space dependence of the diffusion coefficients. In principle, the diffusion coefficient can depend on the medium in which the particle is located, and thus on the concentration of other chemical species at each point in space. For the sake of simplicity, here we consider the diffusion coefficients as constants. We estimated the diffusion constant of ribosomes and polysomes to be  $D_{\smf} = 0.4 \, \mu \rm m^2/s$ and $D_n = 5 \times 10^{-2} \mu \rm m^2/s$, respectively, by leveraging experimental data \cite{bakshi2012superresolution,sanamrad2014single}. For DNA segments it is harder to obtain an estimate based on experimental data. We therefore assume that, because such segments have a linear dimension similar to that of polysomes,  their diffusion coefficient is of the same order of magnitude as that of polysomes,  i.e., $D_{\smdna} = 10^{-2} \, \mu \rm m^2/s$. 

\begin{figure}
\begin{center}
\includegraphics[scale=1.6]{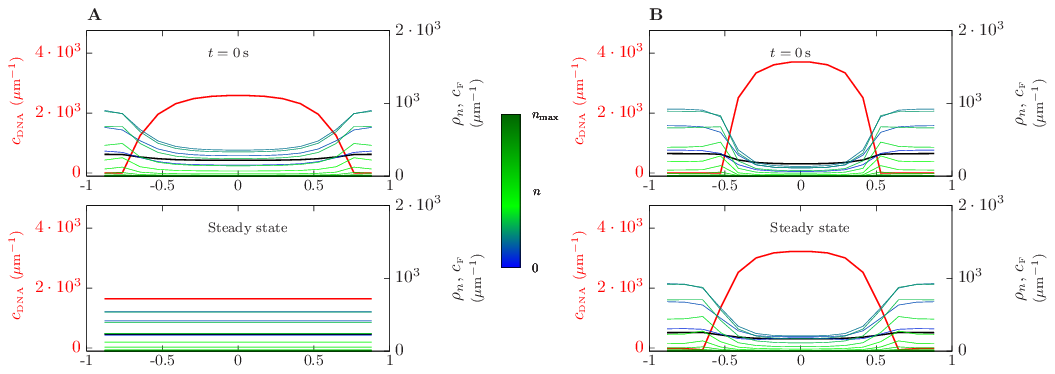}
\figCaption{Effect of third-order virial term}{
\label{3rd_virial} (\textit{A}) Profiles corresponding to the free energy minimization (top) and out-of-equilibrium steady state (bottom), obtained with the second virial coefficients only, for a cell length of $3.6\, \mu {\rm m}$ in the single-chromosome growth scenario. As in the main text, concentrations of DNA, $c_{\smdna}(x)$, and free ribosomes, $c_{\rm F}(x)$  are shown in red and black respectively and polysome concentrations $\rho_n(x)$ are also shown, where the mRNA loading number $n$ is specified by the color box. (\textit{B}) Same quantities as in A, with  the third virial coefficients. }
\end{center}
\end{figure}

Finally, we can evaluate the effect of adding a third virial coefficient in our model. In Fig. \ref{3rd_virial}, we depict the equilibrium minimum of the free-energy and the out-of-equilibrium steady state, both for second and third order in the virial expansion. There is a marked difference, particularly for the out-of-equilibrium steady state.

%\begin{align}
%\label{eq_J}
%J_{\smf}(x,t) = & D_{\smf} \left[ v_{\smf}(x,t) \partial_x c_{\smf}(x,t) - c_{\smf}(x,t) \partial_x v_{\smf}(x,t) \right], \\\label{eq_Jn}
%J_n(x,t) = & D_n \left[ v_n(x,t) \partial_x c_n(x,t) - c_n(x,t) \partial_x v_n(x,t) \right], 
%\end{align} 
%where 
%\be
%v_a(x,t)=e^{-\nu_a(x,t)},
%\ee
%\begin{equation}
%\label{excl_vl}
%\nu_a(x)=\sum_b B_{ab}^{(2)} c_b(x)+\sum_{b,c}B_{abc}^{(3)} c_b(x)c_c(x)\left(1-\frac{\delta_{bc}}{2}\right),
%\end{equation}
%and the sums over $b$ and $c$ run over all species in the system. In \cref{excl_vl}, the term $\delta_{bc}$ is included to avoid double counting when $b=c$.

%By substituting \cref{curr_DNA,eq_J,eq_Jn} in \crefrange{RD_eqs_C}{reac_diff}, we obtain an explicit form for the reaction-diffusion equations. 

\section{Estimates based on the compartment model}\label{estimates}

In this section we will consider the  cell to be divided into three compartments: A central one, the nucleoid, composed exclusively of DNA segments, and two lateral ones which include ribosomes and polysomes. The particles interact through steric interactions, described by the virial coefficients.  Therefore, we consider the following free energy for the particles within the compartments:
\begin{equation}
F_i=F_\textrm{ideal}+k_{{\rm B} T}\frac{N_i^2 B_i}{2V_i},
\end{equation}
where $i$ denotes the compartment and $N_i$ the particle number, $B_i$ the virial coefficient that accounts for the steric interactions among the particles within the compartment, 
%F_ideal is not defined
$F_\textrm{ideal}$ is the free energy of the ideal gas, and $V_i$  the compartment volume. Then, the osmotic pressure exerted by the compartments is
\begin{align}
\nn
P_i=&-\frac{\partial F}{\partial V}\\\label{pressures_SI}
=&\frac{k_B T N_i}{V_i}\left(1+\frac{N_iB_i}{2V_i}\right),
\end{align}
as stated in the main text, see \cref{pressures}. 
 
\subsection{Nucleoid size}\label{nuc_s}
In what follows, we consider \cref{pressures_SI} in the nucleoid and in the pole compartments, and estimate the respective values of the virial coefficients in the nucleoid, $B_{\rm n}$, and at the cell poles, $B_{\rm p}$. 

In the nucleoid, DNA-DNA interactions dominate, yielding a value of $B_{\rm n}\simeq 6.4 \cdot10^{-4} {\mu \rm m}^3$. For the poles we provide an effective value of the virial coefficient by assuming that all ribosomes are bound to mRNAs, and are equally distributed among them, that is, the poles are occupied by spheres all equal in size.  
Given that the ratio of ribosomes to mRNAs changes with the amount of mRNA in the cell, the virial coefficients depend on this last parameter. 
In the cases analyzed in Fig. \ref{fig1}D, we obtain the following values for the virial coefficients, \cref{virial_sph}, for the corresponding mRNA concentrations:
\begin{equation}
B_{\rm p}(1500\, {\mu \rm m}^{-1})= 4.3 \cdot 10^{-4}{\mu \rm m}^3,\quad B_{\rm p}(2400\, {\mu \rm m}^{-1})=3.7 \cdot 10^{-4}{\mu \rm m}^3, \quad B_{\rm p}(3000\, {\mu \rm m}^{-1})= 3.5 \cdot 10^{-4}{\mu \rm m}^3.
\end{equation}

By equating the pressures of the compartments, we obtain the equilibrium value for the volumes of each compartment. For filamentous growth,  where the number of ribosomes, mRNAs, and DNA segments scales linearly with size ($N_i\propto  \ell$), we obtain the solution for the nucleoid size $V_{\rm n}=\varphi V$, where $\varphi$, the fraction of total volume occupied by the nucleoid,  depends on the concentration of mRNAs in the cell:
\begin{equation}
\varphi(1500\, {\mu \rm m}^{-1})=0.7, \quad \varphi(2400\, {\mu \rm m}^{-1})=0.61, \quad \varphi(3000\, {\mu \rm m}^{-1})=0.56.
\end{equation}
As shown in Fig. \ref{fig1}D in the main text, these estimates (gray lines) are in good agreement with the numerical solution of the full reaction-diffusion equations (red points).

\subsection{Nucleoid centering}

The centering of the nucleoid can also be understood in terms of the simplified compartment model above.  

We assume that the mRNA synthesized in the nucleoid can diffuse out of the nucleoid to the lateral compartments. If the nucleoid is not centered in the cell and the synthesized mRNA leaves the nucleoid symmetrically to the left and right, then the mRNA density, and thus the osmotic pressure, will increase in the smaller polar compartment, thus pushing the nucleoid towards the center. As a result, the force that we need to consider is the difference in pressure between the poles times the cross-section $\sigma$ of the cell $F=\sigma(P_{\rm L}-P_{\rm R})$, where `L' and `R' denote the left and right pole, respectively. The dynamical equation for the position of the center of mass of the nucleoid, which we denote by $x_n$, is:
\begin{equation}
\label{dxdt}
\frac{dx_n}{dt}=\frac{D}{k_BT}\sigma(P_{\rm L}-P_{\rm R}), 
\end{equation}
where $D$ is a diffusion constant, not necessarily equal to the diffusion constant of DNA segments. Indeed, $D$ is an effective diffusion coefficient that includes collective effects of DNA segments diffusing together and other biological effects like transertion, see main text \textit{Discussion}.

We can provide a lower bound for the time it takes the nucleoid to center by assuming that the drag is small, i.e. the centering of the nucleoid due to a difference in osmotic pressure is only limited by the synthesis of mRNA. In this limit, the nucleoid moves fast enough to prevent a pressure difference between the poles, that is, a quasi-static approximation of   \cref{dxdt},  $d_t x_{\rm n}=0$, that implies  $P_{\rm L}=P_{\rm R}$. Thus, the centering of the nucleoid is controlled by the rate at which the number of polysomes in the lateral compartments changes. In both compartments, the pressure is set by the concentration of polysomes, whose number is set by the following differential equation:
\begin{equation}
\frac{d \rho_{\rm tot_i}}{dt}=\frac{ \alpha c_\smdna}{2} - \beta \rho_{\rm tot_i},
\end{equation}
whose solution is
\begin{equation}
\rho_{\rm tot_i}=\frac{\alpha\,  c_\smdna}{2 \beta} + C_i e^{-\beta t},
\end{equation}
where $C_i$ is a constant that is set by initial conditions. In the case of the initial condition of Fig. \ref{fig5}, $C_i$ takes the value $C_{\rm L}=+0.2\, \alpha c_\smdna/\beta$ and $C_{\rm R}=-0.2\, \alpha \, c_\smdna/\beta$ for the left and right compartment, respectively, as the initial position of the centre of mass of the nucleoid is located at $+0.2 / \ell$ and the amount of mRNA is directly proportional to the volume of each compartment. Since $P_{\rm L}=P_{\rm R}$ we have $\rho_{\rm tot \, {\rm L}}/V_{\rm L}=\rho_{\rm tot \, {\rm R}}/V_{\rm R}$. Assuming that the nucleoid does not change size during this process, we obtain
\begin{equation}
V_{\rm L}=\frac{\rho_{\rm tot \, {\rm L}}}{\rho_{\rm tot \, {\rm L}}+\rho_{\rm tot \, {\rm R}}}(V-V_{\rm n}),
\end{equation}
where $V$ is the total volume of the cell, and $V_{\rm n}$ the volume of the nucleoid. In the previous relation, the only term that is time-dependent is $\rho_{\rm tot \, {\rm L}}$ since 
$\rho_{\rm tot \, {\rm L}}+\rho_{\rm tot \, {\rm R}}$ is constant in time. Therefore, the position of the nucleoid is set by $V_{\rm L}$, which depends only on $\rho_{\rm tot \, {\rm L}}$. This lower bound on the time for centering is depicted in Fig. \ref{fig5}B.

\section{Experimental Methods}
\label{experiments}
\textit{E. coli} wild type strain NCM3722 was used in the study. To achieve different growth rates, cells were cultured at $37^{\circ}$C in chemostats and in batch. For slow growth rates (0.1 and 0.6 $\textrm{h}^{-1}$), carbon-limiting chemostats with corresponding dilution rates were used, whereas for faster growth, batch cultures with glucose minimal media (0.9 $\textrm{h}^{-1}$) and defined rich media (1.7 $\textrm{h}^{-1}$) were used. The chemostat (Sixfors, HT) volume was $300\, \rm mL $ with oxygen and pH probes to monitor the culture. pH was maintained at $7.2\pm 0.1$ and the aeration rate was set at $4.5 \, \rm l/h$. $40 \, \rm mM$ MOPS media (M2120, Teknova) was used with glucose ($0.4\, \%$, Sigma G8270), ammonia ($9.5\, \rm mM$ NH4Cl, Sigma A9434), and phosphate ($1.32\, \rm mM$ K2HPO4, Sigma P3786) added separately. For the defined rich media, additional Supplement EZ 5X and 10X ACGU Solution (Teknova) were added. In carbon-limiting chemostats, glucose concentration was reduced to $0.08\, \%$. All the sample collection happened after chemostat cultures reached steady state or when batch culture reached OD600 $0.3$.

To measure cell size, $750 \, \mu \rm L$ of culture was fixed with 250  $\mu$L 20\% paraformaldehyde at room temperature for 15 minutes, washed with PBS twice, and stored at $4^{\circ}$C until imaging. 1  $\mu$L of cells were then placed on 1\% low-melting agar pad (Calbiochem) made with PBS and imaged with inverted Nikon90i epifluorescent microscope equipped with a 100 $ \times$ 1.4 NA objective (Nikon) and Hamamazu Orca R2 CCD camera. NIS Elements software (Nikon) was used to automate image acquisition for phase contrast images. Segmentation, quantification of fluorescence intensity, and cell-length measurements were further analyzed in MATLAB using customized programs.

To infer ribosome number per cell, cell number per OD600 and total RNA per OD600 were measured separately. Cell number per OD600 was calculated by serial dilution and plating. To measure total RNA, 1.5 mL of culture was pelleted by centrifugation for 1 min at 13,000 X g. The pellet was frozen on dry ice and the supernatant was taken to measure absorbance at 600 nm for cell loss. The pellet was then washed twice with 0.6M HClO4 and digested with 0.3M KOH for 1 hour at $37^{\circ}$C. The solution was then precipitated with 3M HClO4 and the supernatant was collected. The pellet was re-extracted again with 0.5M HClO4. The supernatants were combined and absorbance measured at 260nm using Tecan Infinite 200 Pro (Tecan Trading AG, Switzerland). Total RNA concentration was determined by multiplying the A260 absorbance with 31 ($\mu$g RNA/mL) as the extinction coefficient.

\section{Parameter estimation from experimental data} \label{parameters_est}

\begin{figure}
\begin{center}
\includegraphics[scale=0.97]{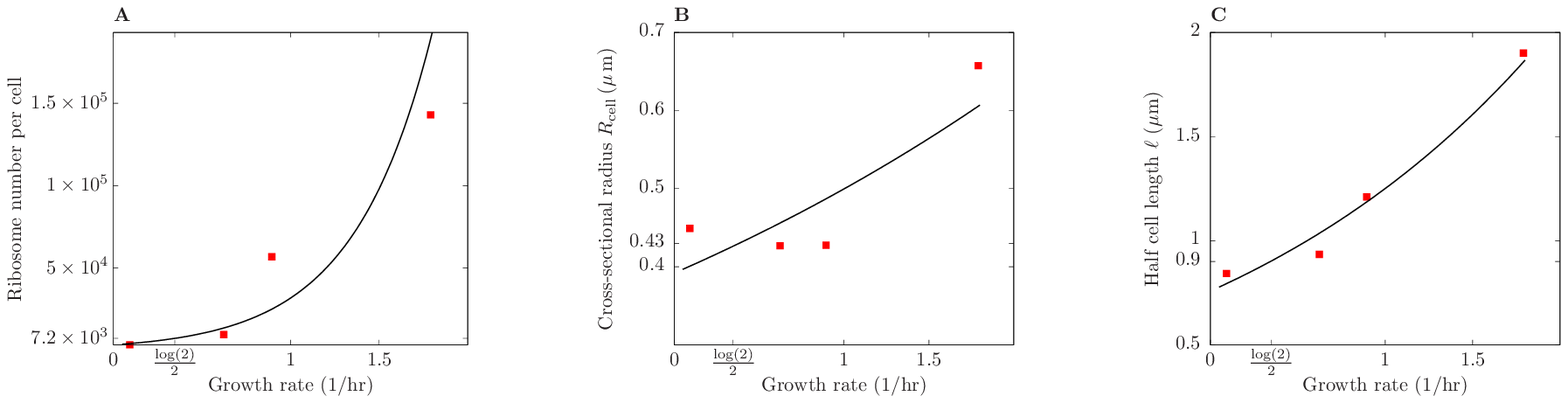}
\figCaption{Experimental data and interpolation}{ \label{exp_data} (\textit{A}) Ribosome number per cell as a function of growth rate. (\textit{B}) Same as \textit{A}, for cell cross-sectional radius. (\textit{C}) Same as \textit{A}, for half cell length. Points correspond to experimental data, and solid curves to the exponential fit. The inferred parameters for the desired growth rate are marked on each axis.}
\end{center}
\end{figure}

The cell  length, cross-sectional radius, and the number of ribosomes, were inferred from experimental data of \textit{E. coli} colonies growing in different chemostatted conditions. The data yields the values of  these parameters for different growth rates. The estimate of the number of ribosomes is derived from the total amount of 23S and 16S ribosomal RNA (rRNA), considering that two-thirds of the total mass of a ribosome comes from rRNA.
The twelve different nutrient limitations analysed correspond to four groups of similar growth rates, and the cell  length and cross-sectional radius, number of ribosomes, and growth rate were averaged across data points belonging to the same growth rate. 
 Finally, the values of the parameters as functions of the growth rate were fitted with an exponential by using the least-squares method. The parameter values for a growth rate of $\log (2)/2 / \rm hr$, which corresponds to a doubling time of $2\, \rm hr$, were obtained via interpolation, by evaluating the exponential fit at the reference growth rate $\log (2)/2 / \rm hr$, see the main text \textit{Model parameters}. The interpolations are shown in Fig. \ref{exp_data}, and the values of the corresponding parameters are given in the main text in \textit{Model parameters}.

\section{Numerical methods}
\label{numerics}
For both the minimization of the free energy \eqref{F_tot} and the time integration of the reaction-diffusion \crefrange{RD_eqs_C}{reac_diff}, numerical methods are required. We use a finite-difference scheme to first order in $\Delta x$ to discretize the spatial degrees of freedom of the system as follows: 
\begin{equation}
2 \ell = \Delta x N_{\rm d},
\end{equation}
where $N_{\rm d}$ is the number of points taken to describe the concentration profile of each of the species in the system, and was taken to be $N_{\rm d}=32,\,64$, depending on the desired accuracy. With this discretization, we evaluated the spatial derivatives in \cref{F_tot}, and obtained a minimum of the free energy by using an algorithm for constrained gradient-based optimization \cite{KraftOptimizationAlgorithm}. We used the C implementation  of the NLopt library \cite{NLOPT}.

For the time integration, we wrote down a set of ordinary differential equations, where to each point in the mesh corresponds a function of time, and such functions are coupled to each other through the local chemical reactions, and to neighboring points in space through the discretized spatial derivatives. To solve this system we used an implementation of the backward differentiation formula (BDF) method in Mathematica \cite{wolfram2014mathematica}.

\section{Scaling of the concentration of chemical species for single-chromosome growth} \label{scaling_single_chromosome}

In the single-chromosome case we  assume that the concentration of mRNA and ribosomes scales linearly with the growth rate and, based on the data of Ref. \cite{hanna20xx}, that the growth rate of \textit{E.coli} decreases linearly with cell length until it reaches zero at $\sim 20\, \mu {\rm m}$. In order to compare the model predictions with the experimental data in  Ref. \cite{wu2019cell}, we assume the same linear law, but with a slope such that the growth rate, $g$,  reaches zero at $30 \, \mu {\rm m}$, because in the data of Ref. \cite{wu2019cell} cells appear to grow up to that length. In addition, we assume that the mRNA degradation rate, $\beta$, decreases linearly in the same way the growth rate does, motivated by the expectation that for slow growth it would be inefficient to turn over mRNA quickly. 

With the above considerations, we can write the following relations:
\begin{align}
g(\ell)&=g_0-a\, \ell ,\\
N_{\rm mRNA}(\ell)&= 2 \ell N_{\rm mRNA_0}g(\ell),\label{mrna_conc} \\
\beta(\ell)&=\beta_0(g_0-a\, \ell),  \label{beta}
\end{align} 
where, for a cell that grows up to $30\,  \mu  \textrm{m}$, $a=g_0/(15\, \mu \textrm{m})$ and the constants $g_0$, $N_{\rm mRNA_0}$, and $\beta_0$ are set by knowing that the values of their respective functions must match those of the reference cell of the main text, see \textit{Model parameters}. Those are:
\begin{align}
g(\ell=0.9\, \mu\textrm{m})&=\frac{\log(2)}{2} \textrm{hr}^{-1}, \\
\frac{N_{\rm mRNA}(\ell=0.9\, \mu\textrm{m})}{2 \ell \sigma}&=2400 \, \mu\textrm{m}^{-3},\\
\beta(\ell=0.9\, \mu\textrm{m})&=3 \times 10^{-3} / \textrm{s}.
\end{align} 
Then, the mRNA synthesis rate, $\alpha(\ell)$, is fixed by the relation $\alpha(\ell) N_{\rm DNA}= \beta (\ell)N_{\rm mRNA}(\ell)$ see the main text \textit{Model parameters}, where  $N_{\rm mRNA}(\ell)$ and $\beta(\ell)$ are given by   \cref{mrna_conc,beta}.

 \begin{figure}
\begin{center}
\includegraphics[scale=1.2]{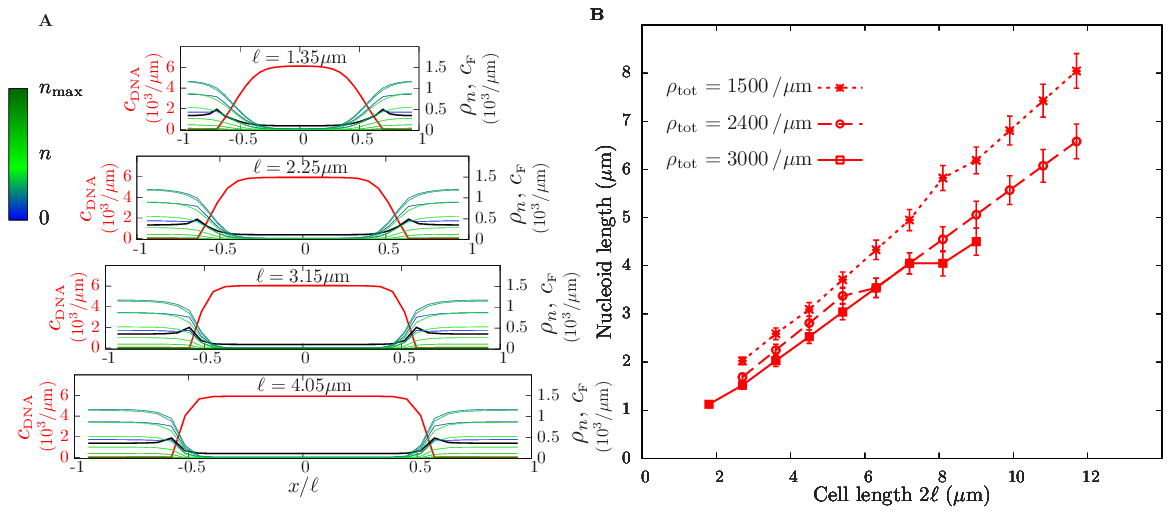}
\figCaption{Equilibrium concentration profiles for \textit{E. coli} growing filamentously}{
\label{equil_const_crowd} (\textit{A}) and (\textit{B}) are the same as Fig. \ref{fig1}C and D in the main text, respectively, but without the out-of-equilibrium terms in the model.}
\end{center}
\end{figure}

\section{Supplementary figures and movies}

\subsection{Figure \ref{equil_const_crowd}}

This figure depicts the results obtained for the filamentous growth scaling as in the main text, section \textit{Filamentous growth}, but in the absence of the out-of-equilibrium chemical reactions, i.e.,  for a passive system. The format of the Figure is the same as that of Fig. \ref{fig1}.

 \begin{figure}
\begin{center}
\includegraphics[scale=1.5]{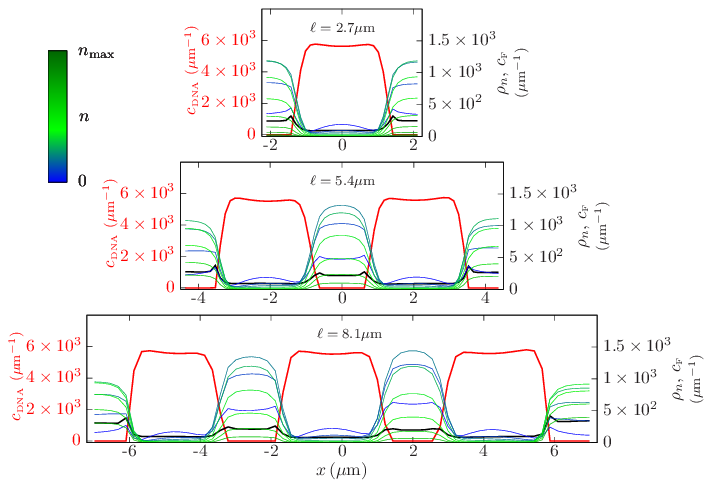}
\figCaption{Steady-state profiles for filamentously growing cells}{
\label{fig_lobes} Steady-state profiles of the concentrations of the components of the \textit{E. coli} 
transcriptional-translational machinery (TTM) for filamentously growing cells with mRNA concentration 
$\rho_{\textrm{tot}}=2400\, \mu \textrm{m}^{-1}$. Colors represent different chemical species of the TTM, as in  Fig. \ref{fig1}C. As in Fig.  \ref{fig1}D,  for cells larger than $\sim 8 \, \mu\textrm{m}$ the nucleoid splits in two lobes. Furthermore, for cells around $\sim 16\,  \mu\textrm{m}$ the nucleoid has three distinct lobes, suggesting  that this may be a pattern with a characteristic length that exists also for longer cells.}
\end{center}
\end{figure}

\subsection{Figure \ref{fig_lobes}}

This figure depicts out-of-equilibrium steady states in the  filamentous-growth case, for cells longer than those shown in the main text. For such long cells (Fig. \ref{fig_lobes} lower panel) there are more than two lobes, suggesting the emergence of a characteristic length scale. Although larger cells could not be examined because the free-energy minimization is too computationally  costly, we expect that more lobes will form in longer cells, with a fixed lobe characteristic size  ($\sim 5\, \rm \mu m$). 

Experimentally, in long filamentously growing cells (with cell division blocked but not DNA replication), nucleoids are observed at tightly controlled positions and distances \cite{Wehrens2018}. However, in Ref. \cite{Wehrens2018} a separate nucleoid appears every $\sim 2.25\, \rm \mu m$ the cell grows in length. This value is roughly half of the one predicted by this model, possibly due to uncertainties in the parameters and lack of modelled connectivity among DNA segments, see \textit{Discussion}.

\subsection{Movie S1}

This movie shows the dynamics of nucleoid splitting, and has been obtained by integrating forward in time  \crefrange{RD_eqs_C}{reac_diff} by using the equilibrium profile as initial condition, see Fig. \ref{fig3}. Concentrations on the $y$ axis are given in units of ${\rm \mu m}^{-1}$, and the $x$ axis is normalized by half the cell length $\ell$. The total polysome concentration is obtained by summing over all polysome species, including bare mRNAs, for each point of space, i.e. $\rho_{\textrm{tot}}(x,t)=\sum_{n=0}^{n_\textrm{max}} \rho_n(x,t)$.

\subsection{Movie S2}

This movie shows the centering dynamics of the nucleoid,  and has been obtained by integrating forward in time  \crefrange{RD_eqs_C}{reac_diff}, see Fig. \ref{fig5}. The initial condition has been obtained by shifting towards one side of the cell  the out-of-equilibrium steady-state profiles. Concentrations on the $y$ axis are in units of ${\rm \mu m}^{-1}$, and the $x$ axis is normalized by half the cell length $\ell$.

\end{document}